\documentstyle[12pt,multibox]{article}
\newcommand{\proof}[1]{\vspace{-.25cm}{\bf Proof:}
#1~$\Box$.\vspace{.30cm}}
\newcommand{\bb}{\begin{equation}}
\newcommand{\ee}{\end{equation}}
\newcommand{\bqn}{\begin{eqnarray}}
\newcommand{\eqn}{\end{eqnarray}}

\newtheorem{theorem}{Theorem}

\begin{document}
\begin{titlepage}

\begin{flushright}

ULB-TH-98/17

ESI 644 (1998)

hep-th/9812140
\end{flushright}

\begin{center} {\Large {\bf The Wess-Zumino Consistency
Condition For
$p$-Form Gauge Theories}}

\end{center}
\vfill

\begin{center} {\large Marc Henneaux$^{a,b}$ and Bernard
Knaepen$^{a
\ *}$}
\end{center}
\vfill

\begin{center}{\sl
$^a$ Physique Th\'eorique et Math\'ematique, Universit\'e
Libre de Bruxelles,\\ Campus Plaine C.P. 231, B--1050
Bruxelles, Belgium\\[1.5ex]

$^b$ Centro de Estudios Cient\'\i ficos de Santiago,\\
Casilla 16443, Santiago 9, Chile\\[1.5ex]

}\end{center}

\vfill

\begin{abstract} The general solution of the
antifield-independent Wess-Zumino consistency condition is
worked out for models involving exterior form gauge fields
of arbitrary degree.  We consider both the free theory and
theories with Chapline-Manton couplings.  Our approach
relies on solving the full set of descent equations by
starting from the last element down (``bottom").
\end{abstract}
\vfill
\vskip 0.2cm
\rule{3cm}{0.1mm}

\noindent
\footnotesize{{\tt * henneaux@ulb.ac.be,
bknaepen@ulb.ac.be}}
\end{titlepage}

\section{Introduction}
\setcounter{equation}{0}
\setcounter{theorem}{0}

$p$-form gauge theories are generalizations of
electromagnetism in which the vector potential -- a 1-form
-- is replaced by exterior forms of higher degree.  They
play an important r\^ole in supergravity and superstring
theory.

The purpose of this paper is to derive the general solution
of the antifield-independent Wess-Zumino consistency
condition \cite{WZ} for free $p$-form theories -- and
theories with interactions that do not deform the gauge
symmetry of the free case -- as well as for theories
involving Chapline-Manton couplings \cite{CM}. As is well
known, the Wess-Zumino consistency condition is central to
any gauge theory.  For instance, at ghost number one, it
constrains the candidate anomalies, while at ghost number
zero, it determines the possible counterterms.

Counterterms and anomalies may actually depend also on the
antifields
\cite{BRS,Zinn,BV,VT,ItzZu,PiguetS,Weinberg}. Thus, one
should in fact solve the Wess-Zumino consistency condition
in the enlarged space containing these variables.  However,
as experience with the Yang-Mills theory indicates
\cite{BBH2}, it is useful to first work out the restricted,
antifield-independent problem before tackling the full
question. The complete solution of the Wess-Zumino
consistency condition in the space of functions of the
fields and the antifields will be given elsewhere \cite{HK}.

The Wess-Zumino consistency condition takes the form
\bb
\gamma a + db = 0.
\label{WZ}
\ee Here, $a$ and $b$ are local spacetime forms (for a
precise definition, see below) while the differential $d$
is the familiar exterior derivative operator acting on
local forms.  The differential
$\gamma$ is the BRST differential in the space of the
fields, the ghosts and the ghosts of ghosts, whose explicit
form depends on the theory at hand.  It will be explicitly
written down in the next section. Trivial solutions of
(\ref{WZ}) read $a = \gamma m + d n$ for some local forms
$m$ and $n$. The physically distinct solutions of the
Wess-Zumino consistency condition are obtained by
quotienting out the trivial ones and are thus parametrized
by the cohomology groups $H(\gamma \vert d)$ of $\gamma$
modulo $d$.   The problem raised by the Wess-Zumino
consistency condition is therefore a cohomological one.

Our method for investigating (\ref{WZ}) follows closely the
approach developed in \cite{DVTV0,DVTV,DVTV1,MT} for
Yang-Mills gauge theories. In that approach, one solves 
(\ref{WZ}) ``from the bottom", i.e., one writes down the
descent equations that follow from the Wess-Zumino
consistency condition and determines the most general
bottom of the descent.  This leads to a generalization of
the ``transgression" formula for $p$-form theories, in
which one associates with each non trivial last ghost of
ghost a gauge-invariant curvature with opposite statistics.

Our main result is that the most general solution of
(\ref{WZ}) can be expressed in terms of exterior products
and exterior derivatives  of the basic form-variables, up
to terms that descend trivially and up to trivial
solutions. 
We say that a solution $a$ descends trivially if it is
equivalent to a strictly $\gamma$-closed solution, i.e.,
$a=a'+\gamma m +dn$, with $\gamma a'=0$.
Thus, it is not
necessary to assume that
$a$ and
$b$ in (\ref{WZ}) depend only on the basic variables
through the exterior forms they define and their exterior
derivatives.  One may allow a priori for a general
dependence on the individual components of the basic
variables and their independent derivatives of arbitrarily
high (but finite) order. The fact that only exterior forms
are relevant (up to solutions that descend trivially)
emerges from the analysis and is not a restriction.  This
result generalizes the theorems established for
$1$-forms in \cite{BDK1,DVHTV} and justifies in particular
the usual methods followed in the literature for
determining the anomalies.

Our paper is organized as follows. In Section 2, we
introduce the various models considered in the article along
with their respective BRST algebra. Then, in Section 3, we
calculate for each model the cohomology $H(\gamma)$ of the
BRST differential. The knowledge of this
cohomology is essential in order to solve to so-called
descent equations, the study of which constitutes the core
of our paper. In Section 4, we explicitely show how the
analysis of the descent proceeds. In Section 5, we prove our
main result, i.e., that any solution of (\ref{WZ}) which
descends non-trivially is an element of the exterior
algebra generated by the form-fields, the ghosts and their
exterior derivative
(up to trivial solutions). Section 6 is then devoted to the
explicit calculation of the auxiliary cohomologies $E_k$
introduced in Section 4, which enable one to obtain the
non-trivial solutions of the Wess-Zumino consistency
condition. Finally, in Section 7, we exhibit the various
counterterms and anomalies for the models defined in
Section 2.

\section{The models}
\setcounter{equation}{0}
\setcounter{theorem}{0}

\subsection{Generalities}

We consider an arbitrary system of exterior form gauge
fields
$B^a_{\mu_1 \ldots \mu_{p_{a}}}$ of degree
$\geq 1$.  Each exterior form gauge field is accompanied by
ghosts and ghosts of ghosts of decreasing form degree and
increasing ghost number
\cite{Siegel,BF,Thierry1,Thierry2,BV,HT},
\begin{equation} C^a_{\mu_1 \ldots
\mu_{p_a-1}},\ldots,C^a_{\mu_1
\ldots
\mu_{p_a-j}},\ldots, C^a.
\label{ghosts}
\end{equation} The ghost number of the ``first" ghosts
$C^a_{\mu_1
\ldots\mu_{p_a-1}}$ and their Grassmann parity are equal to
1.   As one moves from one term to the next one to its
right in  (\ref{ghosts}), the Grassmann parity alternates
and the ghost number increases by one unit up to $p_{a}$.

We denote by ${\cal P}$ the algebra of spacetime forms with
coefficients that are polynomials in the fields, ghosts,
ghosts of ghosts and their derivatives.  Thus, $a$ belongs
to ${\cal P}$ if and only if
\bb a = \frac{1}{q!}\alpha_{\nu_1 \ldots \nu_q} ([B^a_{\mu_1
\ldots \mu_{p_a}}], [C^a_{\mu_1 \ldots
\mu_{p_a-1}}], \ldots, [C^a]) dx^{\nu_1} \wedge \ldots
\wedge dx^{\nu_q}
\label{LocalForms}
\ee where the notation $f([y])$ means that $f$ depends on
$y$ and its successive derivatives up to some finite order,
\bb f = f([y]) \Longleftrightarrow f=f(y, \partial_\mu y,
\ldots,
\partial_{\mu_1 \ldots \mu_k} y).
\ee The form $a$ is allowed to have components of various
form degrees (there is a sum over $q$ in
(\ref{LocalForms})) but has no explicit dependence on the
spacetime coordinates
$x^\mu$ since this is the case usually met in practice
(although such a dependence can be handled by the present
methods). {}From now on, we shall drop explicit reference
to the wedge product in formulas like (\ref{LocalForms}).
It is in the algebra ${\cal P}$ that the cohomological
problem of computing the cohomology of $\gamma$ modulo
$d$ will be analysed.

In both the free and interacting cases, the differential
$\gamma$ is first defined on the undifferentiated
generators $B^a_{\mu_1 \dots
\mu_{p_a}}, C^a_{\mu_2 \ldots \mu_{p_a}}, \dots , C^a$ of
the algebra.  It is then extended to the differentiated
generators by requiring
\bb
\partial_\mu \gamma = \gamma \partial_\mu
\ee which, together with $\gamma (dx^\mu) =0$, implies the
anticommutation relation
\bb
\gamma d + d\gamma =0.
\end{equation} Finally, one extends the differential
$\gamma$ to the whole of ${\cal P}$ by using the Leibnitz
rule,
\begin{equation}
\gamma (ab)=  (\gamma a) b + (-)^{\epsilon_a} a (\gamma b),
\end{equation} where $\epsilon_a$ is the Grassmann parity
of $a$.

\subsection{Free theory}

In the free case, the differential $\gamma$ is defined on
the undifferentiated generators by the equations
\begin{eqnarray}
\gamma B^a_{\mu_1 \dots \mu_{p_a}} &=&
\partial_{[\mu_1}C^a_{\mu_2 \ldots \mu_{p_a}]} ,\\
\gamma C^a_{\mu_1 \ldots \mu_{p_a-k_a}} &=&
\partial_{[\mu_1}C^a_{\mu_2 \ldots \mu_{p_a - k_a}]},\\
\gamma C^a = 0
\end{eqnarray} ($k_a = 1, \ldots, p_a-1$). The field
strengths or ``curvatures" are given by
\bb H^a={1\over (p_{a}+1)!} H^a_{\mu_1 \ldots
\mu_{p_{a}+1}}dx^{\mu_1} \ldots dx^{\mu_{p_{a}+1}} =dB^a,
\label{FieldStrength}
\ee where
\bb B^a={1\over p_{a}!} B^a_{\mu_1 \ldots \mu_{p_{a}}}
dx^{\mu_1}
\ldots dx^{\mu_{p_a}}.
\ee and fulfill
\bb
\gamma (H^a_{\mu_1 \ldots \mu_{p_{a}+1}}) = 0.
\ee

One can rewrite the BRST variations in terms of forms and
exterior derivatives.  This yields
\begin{eqnarray}
\gamma B^a + dC^a_1 &=&0 ,\\
\gamma C^a_1 + dC^a_2 &=&0,\\ &\vdots& \nonumber \\
\gamma C^a_{p_a-1} + dC^a_{p_a} &=&0,\\
\gamma C^a_{p_a} & = &0.
\end{eqnarray} In these equations, $C^a_{j}$ ($j= 1,
\ldots, p_a$) are, up to numerical factors chosen to make
the equations right, the
$(p_a-j)$-forms whose components are respectively
$C^a_{\mu_1 \ldots
\mu_{p_a-j}}$.

It is sometimes convenient to adopt a more uniform notation
that combines the ghosts and the fields, and to indicate
explicitly the form degree and the ghost number. Thus one
sets $B^a \equiv B^{a \, (p_a,0)}$ and $C^a_k \equiv B^{a
\, (p_a - k, k)}$.  In these notations, the BRST equations
are
\bb
\gamma B^{a \, (p_a - k, k)} + d B^{a \, (p_a - k- 1, k+1)}
= 0, \; \;
\gamma B^{a \, (0, p_a)} = 0.
\ee ($k = 0, \dots, p_a - 1$).

Because the gauge transformations of $p$-form gauge
theories whose cou\-plings involve only the curvatures
$H^a_{\mu_1 \dots
\mu_{p_a+1}}$ and their derivatives, or Chern-Simons
couplings
\cite{DJT}, are identical with those of the free theory,
the above BRST transformations and field strength
expressions encompass also these cases.  The ``free
theories" cover therefore a larger class of models.

\subsection{Chapline-Manton models} It has been proved in
\cite{MH,HK0} that the interactions between a set of
exterior form gauge fields are severely constrained by the
consistency requirement that the number of independent
gauge symmetries should be unchanged when the interactions
are switched on.  This result complements the geometric
analysis of \cite{CT} where it was shown that the
non-abelian Yang-Mills construction cannot be generalized to
$p$-forms viewed as connections for extended objects.
[Topological field theory offers ways to bypass some of the
difficulties
\cite{Baulieu}, but will not be discussed here].

Among the few possible consistent interactions, the
Chapline-Manton couplings are particularly interesting
because the gauge algebra remains closed off-shell and the
reducibility identities hold strongly even after the
interactions are switched on.  This is crucial here as it
allows one to investigate the $\gamma$-cohomology without
the antifields.  In general, the new gauge algebra closes
only on-shell and the reducibility identities become
on-shell relations.  This occurs for the celebrated
Freedman-Townsend interaction \cite{FT} and its
generalizations
\cite{HK0} (see also \cite{Dragon}). In that case,
$\gamma^2 \approx 0$ and it is meaningless to consider the
strong cohomology of
$\gamma$ since $\gamma$ is no longer a differential. One
must either work on-shell, or, equivalently, introduce the
antifields to recover nilpotency.  In the Chapline-Manton
models,
$\gamma^2 = 0$ strongly and without the antifields.  One
can thus consider the sub-problem of computing the
cohomology of $\gamma$ in the the algebra ${\cal P}$, which
is an important step in the calculation of the full BRST
cohomology.

The Chapline-Manton interaction has been much studied in
connection with the Green-Schwarz anomaly cancellation
mechanism
\cite{GS,Baulieu2}.  Its construction follows a generic
algebraic pattern discussed in \cite{Baulieu1}.  Rather
than discussing the general case, which would lead to non
informative and uncluttered formulas, we shall consider
four particular examples which illustrate the general
construction.

The Chapline-Manton model is characterized by
gauge-invariant curvatures
$H^a$ which differ from the free ones by terms that vanish
with the coupling constant $g$,
\bb H^a = dB^a + g \mu^a + O(g^2)
\ee The gauge transformations are
\bb
\delta_\epsilon B^a = d \epsilon^a + g \rho^a + O(g^2).
\ee Here, $\mu^a$ is given by a sum of exterior products of
$B$'s and
$dB$'s -- which must match the form degree of $dB^a$ --
while $\rho^a$ is given by a sum of exterior products of
$B$'s, $dB$'s and
$\epsilon$'s (linear in the $\epsilon$'s). The modified
curvatures and gauge transformations must fulfill the
consistency condition
\bb
\delta_\epsilon H^a = 0
\label{invar}
\ee That is, the modified curvatures should be invariant
under the modified gauge transformations. Furthermore,
off-shell reducibility must be preserved, i.e.,
$\delta_\epsilon B^a$ should identically vanish for
$\epsilon^a = d
\lambda^a +  \theta^a $ for some appropriate
$\theta^a(\epsilon, B, dB,g)$. The Lagrangian is a function
of the curvatures and their derivatives, ${\cal L}= {\cal
L}([H^a_{\mu_1 ...
\mu_{p_a+1}}])$ and is thus automatically gauge-invariant.
To completely specify the model, it is thus necessary to
give, besides the field spectrum, the modified curvatures
and gauge transformations fulfilling (\ref{invar}). In many
cases, the curvatures are modified by the addition of
Chern-Simons forms of same degree, but this is not the only
possibility as the example 3 below indicates. We shall set
in the sequel the coupling constant $g$ equal to one.

\subsubsection{Model 1} The first example contains one
$p$-form, denoted by $A \equiv A^{(p,0)}$, and one
$(p+1)$-form, denoted by $B
\equiv B^{(p+1,0)}$. The new field strengths are
\bb F = dA + B, \; H = d B
\label{CM1a}
\ee while the modified gauge transformations take the form
\begin{eqnarray}
\delta_{\epsilon,\eta} A &=& d \epsilon - \eta,
\label{CM1b}\\
\delta_{\epsilon,\eta} B &=& d \eta
\label{CM1c}
\end{eqnarray} where $\epsilon$ is a $(p-1)$-form and
$\eta$ a
$p$-form. The gauge transformations are abelian and remain
reducible off-shell since the choice of gauge parameters
$\epsilon = d \rho +\sigma$,
$\eta = d \sigma$ clearly leads to no variations of the
fields. The BRST transformations of the undifferentiated
generators are
\bb
\gamma A^{(p-k,k)} + d A^{(p-k-1,k+1)} + B^{(p-k,k+1)} = 0
\ee for the $A$-variables, and
\begin{eqnarray}
\gamma B^{(p+1-k,k)} + d B^{(p-k,k+1)} &=& 0, \\
\gamma B^{(0, p+1)} &=& 0
\end{eqnarray} ($k= 0, \dots, p$) for the $B$-ones. One has
\bb
\gamma F = 0 = \gamma H.
\ee This model describes in fact a massive $(p+1)$-form.
Indeed, one can use the gauge freedom of $B$ to set $A
=0$.  Once this is done, one is left with the Lagrangian
for a massive $(p+1)$-form.

\subsubsection{Model 2} The second example contains an
abelian
$1$-form $A \equiv A^{(1,0)}$ and a $2r$-form $B \equiv
B^{(2r,0)}$ ($r>0$). The field strengths are
\bb F = dA , \; H = d B +  F^r A
\label{CM2a}
\ee with $F^r \equiv FF \cdots F$ ($r$ times). The gauge
transformations read
\begin{eqnarray}
\delta_{\epsilon,\eta} A &=&  d\epsilon
\label{CM2b}\\
\delta_{\epsilon,\eta} B &=& d \eta -
 F^r\epsilon
\label{CM2c}
\end{eqnarray} and clearly leave the curvatures invariant.

The BRST transformations are
\begin{eqnarray}
\gamma A^{(1,0)} + d A^{(0,1)} &=& 0, \\
\gamma A^{(0,1)} &=& 0, \\
\gamma B^{(2r,0)} + d B^{(2r -1, 1)} + F^r A^{(0,1)} &=& 0,
\\
\gamma B^{(2r-k,k)} + d B^{(2r -k -1, k+1)} &=& 0, \\
\gamma B^{(0,2r)} &=& 0
\end{eqnarray}
$(k=1, ..., 2r-1)$.

\subsubsection{Model 3} Let $A$, $B$ and $C$ be
respectively 1-, 2- and 3-forms. Define the curvatures
through
\bb F =dA + B, \; H = dB , \;G = dC +  A dB + (1/2)  B^2.
\label{CM3a}
\ee The gauge transformations are
$\delta_{\epsilon,\Lambda,\mu} A = d \epsilon -  \Lambda$,
$\delta_{\epsilon,\Lambda,\mu} B = d \Lambda$ and
$\delta_{\epsilon,\Lambda,\mu} C = d \mu -  \epsilon dB -
 \Lambda B$, where $\epsilon$, $\Lambda$ and $\mu$ are
respectively 0-, 1- and 2-forms.  The gauge algebra is
non-abelian and one easily verifies that the gauge
transformations are off-shell reducible.

The BRST differential is defined by
\begin{eqnarray}
\gamma A^{(1,0)} + dA^{(0,1)} + B^{(1,1)} &=& 0, \\
\gamma A^{(0,1)} + B^{(0,2)} &=& 0, \\
\gamma B^{(2,0)} + d B^{(1,1)} &=& 0, \\
\gamma B^{(1,1)} + d B^{(0,2)} &=& 0, \\
\gamma B^{(0,2)} &=& 0, \\
\gamma C^{(3,0)} + d C^{(2,1)} + A^{(0,1)} H + B^{(1,1)}
B^{(2,0)}&=& 0, \\
\gamma C^{(2,1)}  + d C^{(1,2)} +\frac{1}{2}
B^{(1,1)}B^{(1,1)} + B^{(0,2)}B^{(2,0)}&=& 0, \\
\gamma C^{(1,2)} + d C^{(0,3)} +  B^{(1,1)} B^{(0,2)} &=&
0, \\
\gamma C^{(0,3)} + \frac{1}{2} B^{(0,2)}B^{(0,2)} &=& 0
\end{eqnarray} This example arises in some formulations of
massive supergravity in $10$ dimensions \cite{XX,XXX}.

\subsubsection{Model 4} Our final example is the original
Chapline-Manton model, involving a Yang-Mills connection
$A^a$ and a
$2$-form
$B$.  We assume the gauge group to be $SU(N)$ for
definiteness although the analysis proceeds in the same way
for any other compact group.  The curvatures are
\begin{eqnarray} F &=& dA + A^2 \label{CM4a} \\ H  &=& dB +
\omega_3
\label{CM4b}
\end{eqnarray} where $\omega^{(3,0)} (A, dA)$ is the
Chern-Simons
$3$-form
\bb
\omega_3 = \frac{1}{2} [tr(A dA + \frac{2}{3} A^3)].
\ee The BRST differential reads
\begin{eqnarray}
\gamma A+DC &=& 0, \\
\gamma C+C^2 &=& 0, \\
\gamma B+  \omega_2 + d\eta &=& 0, \\
\gamma\eta +  \omega_1 + d\rho &=& 0, \\
\gamma\rho + \frac{1}{3} tr C^3 &=& 0.
\end{eqnarray} Here, the one-form $\omega_1$ and the
two-form
$\omega_2$ are related to the Chern-Simons form $\omega_3$ 
through the descent,
\begin{eqnarray}
\gamma\omega_3 + d \omega_2&=&0, \ \ \omega_2= tr (CdA), \\
\gamma \omega_2 +d\omega_1&=&0, \ \ \omega_1= tr (C^2 A),\\
\gamma \omega_1 + d(\frac{1}{3} tr C^3)&=&0.
\end{eqnarray} The full BRST cohomology of this model was
worked out in
\cite{HK2}, so we shall only illustrate here some aspects
of the descent equation that were not explicitly discussed
in that paper.

\section{BRST cohomology $H^*(\gamma)$}
\setcounter{equation}{0}
\setcounter{theorem}{0}

\subsection{Free Theory} In order to compute the general
solution of the Wess-Zumino consistency condition, one
needs to know $H(\gamma)$, i.e., the general solution of
$\gamma a = 0$ modulo coboundaries ($a=\gamma b$). We start
with the free models.  The cohomology
$H(\gamma)$ for this case was given in \cite{HKS1} but
without giving all the details. This is done here.
\begin{theorem} The cohomology of $\gamma$ is given by,
\begin{equation} H(\gamma) = {\cal I} \otimes {\cal C},
\end{equation} where ${\cal C}$ is the algebra generated by
the ``last", {\em undifferentiated} ghosts of ghosts
$C^a_{p_a}$ ($\equiv B^{a (0,p_a)}$), and ${\cal I}$ is
the algebra generated by the fields strength components
$H^a_{\mu_1 \dots \mu_{p_a +1}}$ and their derivatives.
\end{theorem}
\proof{One follows the standard method which consists in
separating the variables into three sets obeying
respectively $\gamma x^i=0,\
\gamma y^\alpha=z^\alpha$,
$\gamma z^\alpha=0$. The variables $y^\alpha$ and
$z^\alpha$ form ``contractible pairs" and the cohomology is
generated by the (independent) variables $x^i$
\cite{Sullivan}. In our case, the $x^i$ are given by the
fields strength components, their derivatives and the last
(undifferentiated) ghosts of ghosts.

To arrive at the appropriate decomposition, we split the
generators of ${\cal P}$, which are the successive
derivatives
$\partial_{\alpha_1\ldots\alpha_k} B^{a (l,
p_a-l)}_{\mu_1\ldots
\mu_l}$ of the fields and the ghosts, into irreducible
tensors under the full linear group
$GL(n)$. Since the
$\partial_{\alpha_1\ldots\alpha_k}B^{a (l,
p_a-l)}_{\mu_1\ldots
\mu_l}$ are completely symmetric in $\alpha_1\ldots
\alpha_k$ and completely antisymmetric in $\mu_1\ldots
\mu_l$, they transform under
$GL(n)$ as the variables of the tensor product
representation symbolically denoted by

\begin{picture}(0,105)(0,0)
\multiframe(0,50)(15.5,0){5}(15,15){$\alpha_1$}...{$\alpha_k$}
\put(85,55){$\bigotimes$}
\multiframe(105,20)(0,15.5){5}(15,15){$\mu_l$}...{$\mu_1$}
\put(135,55){$\simeq$}
\multiframe(170.5,82)(15.5,0){5}(15,15){$\alpha_1$}...{$\alpha_k$}
\multiframe(155,20)(0,15.5){5}(15,15){$\mu_l$}...{$\mu_1$}
\put(255,55){$\bigoplus$}
\multiframe(290.5,82.5)(15.5,0){4}(15,15){$\alpha_2$}..{$\alpha_k$}
\multiframe(275,5)(0,15.5){6}(15,15){$\alpha_1$}{$\mu_l$}...{$\mu_1$}
\put(215,45){1)}
\put(330,45){2)}
\end{picture}

\noindent in \cite{hamermesh}.

Convenient generators for the irreducible spaces
corresponding to diagrams 1) and 2) are respectively,
\begin{equation}
\partial_{( {\alpha_1\ldots\alpha_k}}B^{a
(l,p_a-l)}_{[\mu_1{)}_1\ldots
\mu_l{]}_2}\hbox{\ \ \ and \ \ \ }
\partial_ {\alpha_2\ldots\alpha_{k}}H^{a (l+1, p_a
-l)}_{\alpha_1\mu_1\ldots
\mu_l},
\end{equation} with $H^{a (l, p_a -l+1)}_{\mu_1\ldots
\mu_l}=\partial_{[\mu_1}B^{a (l-1, p_a - l
+1)}_{\mu_2\ldots\mu_l]}$. In particular, $H^{a (p_a +1,
0)}_{\mu_1\ldots \mu_{p_a +1}}
\equiv H^a_{\mu_1\ldots \mu_{p_a +1}}$.  Here, $[\ ]$ and
$(\ )$ mean respectively antisymmetrization and
symmetrization; the subscript indicates the order in which
the operations are done.  The generators
$\partial_ {\alpha_2\ldots\alpha_{k}}H^{a (l+1, p_a
-l)}_{\alpha_1\mu_1\ldots
\mu_l}$ are obtained by first symmetrizing
$\partial_{\alpha_1\ldots\alpha_k} B^{a (l,
p_a-l)}_{\mu_1\ldots
\mu_l}$ with respect to $\mu_1$,
$\alpha_2$, ..., $\alpha_k$ and then antisymmetrizing with
respect to the $\mu$'s and $\alpha_1$.  In this last step,
the terms with
$\partial_{\mu_1} \partial_{\alpha_1}$ yield zero.

A direct calculation shows that
\begin{eqnarray}
\gamma B^{a (l,p_a-l)}_{\mu_1\ldots\mu_l} &=&H^{a (l, p_a-l
+1)}_{\mu_1\ldots\mu_l}, \\
\gamma
\partial_{( {\alpha_1\ldots\alpha_k}}B^{a
(l,p_a-l)}_{[\mu_1{)}_1\ldots
\mu_l{]}_2}&=&c\partial_{\alpha_1\ldots\alpha_{k}} H^{a (l,
p_a-l +1)}_{\mu_1\ldots
\mu_l}
\end{eqnarray} with $c=\frac{k+l}{l(k+1)}$ and $2 \leq
l\leq p$, and
\begin{eqnarray}
\gamma H^{a (p_a+1, 0)}_{\mu_1\ldots\mu_{p_a+1}}&=&0,\\
\gamma
\partial_{\alpha_1\ldots\alpha_{k}} H^{a (p_a+1,
0)}_{\mu_1\ldots
\mu_{p_a+1}}&=&0.
\end{eqnarray} Similarly, the relations involving the last
ghosts of ghosts are
\begin{eqnarray}
\gamma B^{a(1,p_a-1)}_{\mu_1}&=&\partial_{\mu_1}B^{a
(0,p_a)},
\\\gamma \partial_{(
{\alpha_1\ldots\alpha_k}}B^{a(1,p_a-1)}_{\mu_1{)}}&=&\partial_{\alpha_1
\ldots\alpha_k\mu_1}B^{a (0,p_a)}, \\
\gamma B^{a (0,p_a)}&=&0.
\end{eqnarray} All the generators are are now split
according to the rule recalled at the beginning of the
subsection: the $z$'s are the
$B^{a (l,p_a-l)}_{\mu_1\ldots\mu_l}$ with $1 \leq l\leq p$,
and their derivatives $\partial_{(
{\alpha_1\ldots\alpha_k}}B^{a (l,p_a-l)}_{[\mu_1{)}_1\ldots
\mu_l{]}_2}$ corresponding to the diagram of Young type 1).
The $y$'s are their $\gamma$-variations, i.e., the
$H^{a (l, p_a-l +1)}_{\mu_1\ldots\mu_l}$ and their
derivatives ($2
\leq l\leq p$), and the derivatives
$\partial_{\alpha_1 \ldots\alpha_k\mu_1}B^{a (0,p_a)}$ of
the last ghosts of ghosts. Of course, any identity verified
by the $y$'s is also verified by the $z$'s since they
correspond to identical Young tableaux and hence transform
in the same representation of $GL(n)$.  There are thus as
many independent $y$'s as there are independent
$z$'s.  Finally, the $x$' are the ``left-over" variables,
i.e., the curvatures $H^{a (p_a+1,
0)}_{\mu_1\ldots\mu_{p_a+1}}
\equiv H^{a}_{\mu_1\ldots\mu_{p_a+1}}$, their successive
derivatives and the last, undifferentiated ghosts of ghosts
$B^{a (0,p_a)}$.

The cohomology is therefore generated by
\begin{equation} B^{a (0,p_a)}, \ H^{a (p_a+1,
0)}_{\mu_1\ldots\mu_{p+1}}
\hbox{\ \ and\ \ }
\partial_{\alpha_1\ldots\alpha_{k}} H^{a (p_a+1,
0)}_{\mu_1\ldots
\mu_{p+1}}.
\end{equation} This ends the proof of the theorem. As shown
recently
\cite{GarciaK}, the same method applies to the calculation
of the
$\gamma$-cohomology of gauge fields with more general
symmetry structure \cite{Curtright}. Note that the
generators are not independent but restricted by the
Bianchi identity $dH^{a (p_a+1, 0)}=0$. }

\subsection{Chapline-Manton model 1}

The cohomology $H^*(\gamma)$ of the Chapline-Manton model
can be worked out as in the free case, by exhibiting
explicitly the contractible part of the algebra.  This
contractible part typically gets larger with the coupling:
some cocycles are removed from
$H^*(\gamma)$. This happens for the examples 1, 3 and 4.

In the absence of couplings, the $\gamma$-cohomology for
the first model would be given, as the previous subsection
indicates, by polynomials \\ $P([F_{\mu_1 ...\mu_{p+1}}],
[H_{\mu_1 ...
\mu_{p+2}}], A^{(0,p)}, B^{(0,p+1)})$ in the components of
the curvatures and their derivatives as well as in the last,
undifferentiated ghosts. When the coupling is turned on,
however, some of these ``$x$"-variables become contractible
pairs and get cancelled in cohomology. Specifically, it is
the last ghosts of ghosts that disappear.  To see this, one
first makes the same redefinitions of generators for the
$B$-sector as in the free case. The $x$, $y$ and
$z$-variables are taken as before, except that the last,
undifferentiated ghost of ghost $B^{(0,p+1)}$ counts now as
a
$y$ (see below).

In the $A$-sector, one takes for the $x$'s the improved
field strength components $F_{\mu_1 ...\mu_{p+1}}$ and
their derivatives.  The
$z$-variables are taken as before, i.e., are the
derivatives of
$A^{(l,p-l)}_{\mu_1 ...\mu_l}$ ($l>0$) of Young
symmetry-type 1) and the $y$-variables are just defined to
be their
$\gamma$-variations.  This is an invertible change of
generators provided one includes as well the last,
undifferentiated ghost of ghost
$A^{(0,p)}$, as in the free case.  But this variable counts
now as a
$z$ since it is no longer closed.  And it actually ``kills"
$B^{(0,p+1)}$ since $\gamma A^{(0,p)} + B^{(0,p+1)}=0$.
Thus, the variables $A^{(0,p)}$ and $B^{(0,p+1)}$, which
were previously
$x$-variables, form now a contractible pair and disappear as
announced.

The Bianchi identities for the new field strengths read
\bb dF = H, \; \; dH=0.
\ee They can be used to express the $H$-components and their
derivatives in terms of the components $F_{\mu_1 ...
\mu_{p+1}}$ and their derivatives, which thus completely
generate the cohomology.

To summarize, the $\gamma$ cohomohology is given by
\begin{theorem}
\label{gammaCohoCM1} For the Chapline-Manton model
(\ref{CM1a}), (\ref{CM1b}), (\ref{CM1c}), the cohomology
$H(\gamma)$ is given by the polynomials in the improved
field strength components
$F_{\mu_1 ...\mu_{p+1}}$ and their derivatives,
\bb
\gamma \omega = 0 \Leftrightarrow \omega = \frac{1}{q!}
\omega_{\nu_1 ... \nu_q}([F_{\mu_1 ... \mu_{p+1}}])
dx^{\nu_1} \dots dx^{\nu_q}.
\ee In particular, there is {\em no} cohomology at
non-vanishing ghost number.
\end{theorem}

The situation is very similar to the discussion of the
gauged principal $U(1)$ sigma model
\cite{wilch} (see also \cite{Saliu} in this context).

\subsection{Chapline-Manton model 2} In this case, the
$\gamma$-cohomology is unchanged compared with the free
case (in terms of the improved, gauge-invariant
curvatures).  The last ghosts remain in cohomology because
$A^{(0,1)}$ is still $\gamma$-closed, so the mechanism of
the previous subsection is not operative.  We leave it to
the reader to make the change of generators necessary to
prove the following theorem.
\begin{theorem} The cohomology of $\gamma$ for the
Chapline-Manton model (\ref{CM2a}), (\ref{CM2b}),
(\ref{CM2c}), is given by
\bb H(\gamma)= \tilde{{\cal I}} \otimes {\cal C}
\ee where ${\cal C}$ is the algebra generated by the last,
undifferentiated ghosts $A^{(0,1)}$ and $B^{(0,2r)}$, and
where
$\tilde{{\cal I}}$ is the agebra generated by the gauge
invariant field strength components
$F_{\mu \nu}$, $H_{\mu_1 \dots \mu_{2r+1}}$ and their
derivatives.
\end{theorem} Note the new form of the Bianchi identities
on the curvatures
\bb dF=0,\; \; dH=F^{r+1}.
\ee

\subsection{Chapline-Manton model 3} The discussion of the
third example proceeds to a large extent like that of the
first one.  The last ghosts of ghosts $A^{(0,1)}$ and
$B^{(0,2)}$ form a contractible pair and disappear in
cohomology; the improved last ghost of ghost
\bb
\tilde{C}^{(0,3)}=C^{(0,3)}-\frac{1}{2} A^{(0,1)} B^{(0,2)}
\label{improvedGhost}
\ee remains.  Thus one has
\begin{theorem} The cohomology of $\gamma$ for the
Chapline-Manton model (\ref{CM3a}) is given by
\bb H(\gamma) = \tilde{\tilde{{\cal I}}} \otimes
\tilde{{\cal C}}
\ee where $\tilde{\tilde{{\cal I}}}$ is the algebra
generated by the gauge invariant field strength components
$F_{\mu \nu}$, $G_{\mu \nu
\rho
\sigma}$ and their derivatives, and where $\tilde{{\cal
C}}$ is the algebra generated by the last, improved ghost
of ghost
$\tilde{C}^{(0,3)}
\equiv C^{(0,3)} - \frac{1}{2}A^{(0,1)}B^{(0,2)}$.
\end{theorem} Again, note the new form of the Bianchi
identities,
\bb dF = -H, \; dH = 0, \; dG = -F H
\label{Bianchi5}
\ee which enable one in particular to express $H$ in terms
of the derivatives of $F$.

\subsection{Chapline-Manton model 4} The
$\gamma$-cohomology for this model is explicitly given in
\cite{HK2}.  It is given by the theorem:
\begin{theorem} The cohomology of $\gamma$ for the
Chapline-Manton model (\ref{CM4a}), (\ref{CM4b}) is given by
\bb H(\gamma) = {\cal J} \otimes {\cal D}
\ee where (i) ${\cal J}$ is the algebra of the invariant
polynomials in the Yang-Mills curvature components and
their covariant derivatives, as well as in the components
of the gauge invariant curvature $H$ and their derivatives;
and (ii) ${\cal D}$ is the algebra generated by the
``primitive forms" $trC^5$, $trC^7$, ...,
$trC^{2N-1}$.
\end{theorem} We recall that the Lie algebra cohomology for
$SU(N)$ is generated by the primitive forms $trC^3$,
$trC^5$, ... up to
$trC^{2N-1}$
\cite{Greub,Koszul}.  The primitive form $trC^3$ is removed
from the cohomology of $\gamma$ because it is exact when
the coupling to the
$2$-form is introduced: the last ghost of ghost $\rho$
kills $trC^3$ in cohomology.  The Bianchi identity reads
\bb DF = 0, \; \; dH=trF^2.
\ee

\section{Descent equation and lifts of $\gamma$-cocycles}
\setcounter{equation}{0}
\setcounter{theorem}{0}

\subsection{The descent equation} Our method for solving the
Wess-Zumino consistency condition is that developped in
\cite{DVTV0,DVTV,MT} for the Yang-Mills case.  In that
approach, one analyses the $\gamma$-cocycles that can
appear as ``bottoms" of the descent equations
\cite{descent}.  We shall first briefly summarize the main
ideas.  These depend only on the generic properties of
$\gamma$ and $d$ and not on their specific forms.  We shall
then apply them to the models at hand.

To a given solution $a^{(p,q)}$ of the Wess-Zumino
consistency condition
\bb
\gamma a^{(p,q)} + d a^{(p-1,q+1)} = 0
\ee (where $p$ and $q$ denote respectively the form degree
and the ghost number), one can associate another solution
of the Wess-Zumino consistency condition, namely,
$a^{(p-1,q+1)}$.  Indeed, the triviality of
$d$ (``algebraic Poincar\'e lemma" \cite{trivialityofd})
implies\footnote{In the space of $x$-independent forms, the
cohomology of $d$ is actually not trivial, even in form
degree $<n$ (the case relevant to the descent).  Indeed,
the constant forms are in the cohomology.  But these never
get in the way because $\gamma a^{(p-1,q+1)}$ and the
successive terms in the descent have strictly positive
ghost number.}
\bb
\gamma a^{(p-1,q+1)} + d a^{(p-2,q+2)} =0
\ee for some $a^{(p-2,q+2)}$.  There are ambiguities in the
choice of
$a^{(p-1,q+1)}$ given the class $[a^{(p,q)}]$ of
$a^{(p,q)}$ in
$H^{(p,q)}(\gamma \vert d)$, but it is easy to verify that
the map
$\partial : H^{(p,q)}(\gamma \vert d) \rightarrow
H^{(p-1,q+1)}(\gamma \vert d)$ is well defined.

The map $\partial$ is in general not injective.  There are
non trivial classes of $H^{*,*}(\gamma \vert d)$ that are
mapped on zero through the descent.  For instance, if one
iterates $\partial$, one gets from
$a^{(p,q)}$ a chain of cocycles in $H^{*,*}(\gamma \vert
d)$,
$[a^{(p,q)}] \mapsto \partial [a^{(p,q)}] \in
H^{(p-1,q+1)}(\gamma
\vert d)
\mapsto \partial^2 [a^{(p,q)}] \in H^{(p-2,q+2)} \mapsto
\cdots \mapsto \partial^k [a^{(p,q)}] \in H^{(p-k,q+k)}
\mapsto 0$ which must eventually end on zero since there
are no forms of negative form degree.  The equations
defining the successive images of $[a^{(p,q)}]$ are
\begin{eqnarray}
\gamma a^{(p,q)} + d a^{(p-1,q+1)} &=& 0 \\
\gamma a^{(p-1,q+1)} + d a^{(p-2,q+2)} &=& 0 \\ &\vdots&
\nonumber \\
\gamma a^{(p-k,q+k)} + da^{(p-k-1,q+k+1)} &=& 0  \\
\,  [a^{(p-k-1,q+k+1)}] &=& 0
\end{eqnarray} and are known as the ``descent equations"
\cite{descent}.  Since
$a^{(p-k-1,q+k+1)}$ is trivial in
$H^{(p-k-1,q+k+1)}(\gamma \vert d)$, i.e.,
$a^{(p-k-1,q+k+1)}=
\gamma  b^{(p-k-1,q+k)} +  db^{p-k,q+k+1)}$,   one may
redefine
$a^{(p-k,q+k)}
\rightarrow a^{(p-k,q+k)} - db^{(p-k-1,q+k)} =
a'^{(p-k,q+k)}$ in such a way that we have $\gamma
a'^{(p-k,q+k)} = 0$, i.e.,
$a'^{(p-k-1,q+k+1)}=0$.  Conversely, if $a^{(p-k,q+k)}$ is
annihilated by $\gamma$, then $\partial [a^{(p-k,q+k)}]
=0$.  Thus, the last non-trivial element $a^{(p-k,q+k)}$,
or ``bottom", of the descent is a $\gamma$-cocycle that is
not exact in
$H^{*,*}(\gamma \vert d)$.  The non-injectivity of
$\partial$ follows precisely from the existence of such
cocycles.

The length of the descent associated with $[a^{(p,q)}]$ is
the integer $k$ for which $\partial^k [a^{(p,q)}]$ is the
last non-trivial cocycle occuring in the chain.  One says
that a descent is non trivial if it has length $\geq 1$.
The idea of
\cite{DVTV0,DVTV} is to classify the elements of
$H^{*,*}(\gamma \vert d)$ according to the length of the
associated descent.

In order to achieve this, one must determine the possible
bottoms, i.e., the elements of $H(\gamma)$ which are not
trivial in
$H(\gamma \vert d)$ and which can be lifted $k$ times.

\subsection{Lifts of elements of $H(\gamma)$ - An example}

The difficulty in the analysis of the lift is that contrary
to the descent which carries no ambiguity in cohomology,
the lift is ambiguous because
$H(\gamma)$ is not trivial.  Furthermore, for the same
reason, the lift can be obstructed, i.e., given $a \in
H(\gamma)$, there may be no descent (i) which has
$a$ as bottom; and (ii) which starts with a solution
$b$ of the Wess-Zumino consistency condition such that $db
=0$ (while any descent ends always with an $a$ such that
$\gamma a = 0$). The ``first" $b$ may be such that $db
\not= 0$ or even $db \not=
\gamma$(something). In that case, there is no element $c$
above $b$ such that $\gamma c + db =0$ (while there is
always an element $e$ below $a$ such that $\gamma a + de =
0$, namely $e=0$: the descent effectively stops at $a$ but
is not obstructed at $a$).

In this subsection, we shall illustrate these features on a
specific example: that of a free $1$-form $A$ and a free
$2$-form $B$, with BRST algebra
\begin{eqnarray}
\gamma A^{(1,0)} + d A^{(0,1)} &=& 0, \; \gamma A^{(0,1)} =
0 \\
\gamma B^{(2,0)} + d B^{(1,1)} &=& 0, \; \gamma B^{(1,1)} +
d B^{(0,2)} = 0, \; \gamma B^{(0,2)} = 0.
\end{eqnarray} The curvatures are $F= dA$ and $H=dB$, with
$\gamma F=
\gamma H =0$.

Consider the $\gamma$-cocycle $A^{(0,1)}B^{(0,2)}$.  It has
form-degree zero and ghost number three.  The descent that
ends on this bottom has length one, and not the maximum
length three. Indeed, the $\gamma$-cocycle
$A^{(0,1)}B^{(0,2)}$ can be lifted once, since there exists
$a \in {\cal P}$ such that $\gamma a +
d(A^{(0,1)}B^{(0,2)}) = 0$.  One may take $a = A^{(1,0)}
B^{(0,2)} +A^{(0,1)} B^{(1,1)}$. Of course, $a$ has
form-degree one and ghost number two.  If one tries to lift
the given $\gamma$-cocycle
$A^{(0,1)}B^{(0,2)}$ once more, one meets an obstruction. 
Namely, there is no $b$ such that $\gamma b + da = 0$. This
is because $da$ is in the same $\gamma$-class as $F
B^{(0,2)}$, which is non-trivial, i.e., which cannot be
written as a $\gamma$-variation. It is easy to verify that
one cannot remove the obstruction by adding to $a$ a
$\gamma$-cocycle (which would not change
$\gamma a$).  This provides an example of a
$\gamma$-cocycle for which the lift is obstructed after one
step.

Consider now the $\gamma$-cocycle $\frac{1}{2} F
(B^{(0,2)})^2$ with ghost number four and form-degree two. 
This cocycle can be lifted a first time, for instance $F
B^{(1,1)} B^{(0,2)}$ is above it,
\bb
\gamma [F B^{(1,1)} B^{(0,2)}] + d [\frac{1}{2} F
(B^{(0,2)})^2] = 0.
\label{firstliftof}
\ee It can be lifted a second time to  $ \frac{1}{2}
(B^{(1,1)})^2 + F B^{(2,0)} B^{(0,2)}$. However, if one
tries to lift it once more, one meets apparently the
obstruction $F H B^{(0,2)}$, since the exterior derivative
of
$ \frac{1}{2} (B^{(1,1)})^2 + F B^{(2,0)} B^{(0,2)}$
differs from the
$\gamma$-cocycle $F H B^{(0,2)}$ by a $\gamma$-exact term.
It is true that $F H B^{(0,2)}$ is a non-trivial
$\gamma$-cocycle. However, the obstruction to lifting three
times $\frac{1}{2} F (B^{(0,2)})^2$ is really absent.  What
happens is that we made a ``wrong" choice for the term above
$\gamma$-cocycle $\frac{1}{2} F (B^{(0,2)})^2$ and should
have taken not $F B^{(1,1)} B^{(0,2)}$, but rather, a term
that differs from it by an appropriate $\gamma$-cocycle. 
This is because $F H B^{(0,2)}$ is in fact the true
obstruction to lifting twice the $\gamma$-cocycle
$A^{(0,1)} H B^{(0,2)}$.  Thus if one replaces
(\ref{firstliftof}) by
\bb
\gamma [F B^{(1,1)} B^{(0,2)} - A^{(0,1)} H B^{(0,2)}] +
d[\frac{1}{2} F (B^{(0,2)})^2]= 0,
\ee which is permissible since $\gamma(A^{(0,1)} H
B^{(0,2)}) = 0$, one removes the obstruction to lifting
further $\frac{1}{2} F (B^{(0,2)})^2$.  This shows that the
obstructions to lifting $k$ times a $\gamma$-cocycle are
not given by elements of $H(\gamma)$, but rather, by
elements of $H(\gamma)$ that are not themselves
obstructions of shorther lifts.  The ambiguity in the
choice of the lifts plays accordingly a crucial r\^ ole in
the analysis of the obstructions.

In fact, the given $\gamma$-cocycle
$\frac{1}{2} F (B^{(0,2)})^2$ is actually trivial in
$H(\gamma \vert d)$
\begin{eqnarray}
\frac{1}{2} F (B^{(0,2)})^2 =  & & d[\frac{1}{2} A^{(1,0)}
(B^{(0,2)})^2 + A^{(0,1)} B^{(1,1)} B^{(0,2)}] \nonumber \\
& & +
\gamma[A^{(1,0)} B^{(1,1)} B^{(0,2)} + \frac{1}{2}
A^{(0,1)}(B^{(1,1)})^2 + \nonumber \\ & & \; \;
A^{(0,1)}B^{(2,0)} B^{(0,2)}]
\end{eqnarray} and therefore, its lift can certainly never
be obstructed.

\subsection{Lifts of elements of $H(\gamma)$ - The first
two steps}

To control the features that we have just illustrated, it is
necessary to introduce new differential algebras
\cite{DVTV0,DVTV}.  Let $E_0 \equiv H(\gamma)$. We define a
map $d_0: E_0 \rightarrow E_0$ as follows:
\bb d_0[a] = [da]
\ee where $[]$ is here the class in $H(\gamma)$.  This map
is well defined because $\gamma da = -d \gamma a = 0$ (so
$da$ is a
$\gamma$-cocycle) and $d(\gamma m) = -\gamma (dm)$ (so $d$
maps a
$\gamma$-coboundary on a $\gamma$-coboundary). Now, $d_0$
is a derivation and $d_0^2=0$, so it is a differential.
Cocycles of $d_0$ are elements of $H(\gamma)$ that can be
lifted at least once since
$d_0[a] = 0 \Leftrightarrow da+ \gamma b =0$ for some $b$,
so $b$ descends on $a$.  By contrast, if
$d_0[a] \not= 0$, then $a$ cannot be lifted and, in
particular, $a$ is not exact in $H(\gamma \vert d)$ (if it
were, $a = \gamma m + dn$, one would have $da = - \gamma
dm$, i.e., $da = 0$ in $H(\gamma)$).  Let $F_0$ be a
subspace of $E_0$ supplementary to $Ker \, d_0$.  One has
the isomorphism (as vector spaces)
\bb E_0 \simeq Ker \, d_0 \oplus F_0.
\label{iso1}
\ee

The next step is to investigate cocycles that can be lifted
at least twice. In order to be liftable at least once,
these must be in $Ker
\, d_0$ . Among the elements of $Ker \, d_0$, those that
are in $Im
\, d_0$ are not interesting, because they are elements of
$H(\gamma)$ that are trivial in $H(\gamma \vert d)$ ($[a] =
d_0[b] \Leftrightarrow a = db + \gamma m$).  Thus the
relevant space is $E_1 \equiv H(d_0, E_0)$.  One has
\bb Ker \, d_0 \simeq Im \, d_0 \oplus E_1
\label{iso1'}
\ee One then defines the differential $d_1: E_1 \rightarrow
E_1$
\bb d_1 [[a]] = [[db]]
\label{defd1}
\ee where $b$ is defined through $da + \gamma b = 0$ --
recall that
$d_0[a] = 0$ -- and where $[[a]]$ is the class of $[a]$ in
$E_1$.  It is easy  to see that (\ref{defd1}) provides a
well-defined differential in $E_1$\footnote{Proof: $d_0[a]
= 0 \Rightarrow da+\gamma b= 0 \Rightarrow \gamma db=0$. 
Hence, $db$ is a
$\gamma$-cocycle, which is clearly annihilated by $d_0$,
$d_0 [db]= [d^2 b] = 0$. Furthermore, the class of $db$ in
$E_1$ does not depend on the ambiguity in the definition of
$b$, since if $b$ is replaced by $b+dm+ u$ with $\gamma
u=0$, then $db$ is replaced by $db + du$ which is
equivalent to $db$ in
$E_1$ (the class of $du$ in $E_0$ is equal to $d_0[u]$
since $\gamma u = 0$, and this is zero in $E_1$). The
derivation property is also easily verified, $d_1 (ab) =
(d_1a) b + (-1)^{\epsilon_a}a d_1 b$.}.

If $[[a]] \in E_1$ is a $d_1$-cocycle, then it can be
lifted at least twice since $[[db]]=0$ in $E_1$ means $db
=d u + \gamma$(something) with
$\gamma u = 0$.  Thus one has $da + \gamma b' = 0$ with
$b' = b - u$ and $d b' = \gamma$(something). If on the
contrary, $d_1 [[a]] \not= 0$, then the corresponding
elements in $H(\gamma)$ cannot be lifted twice, $d_1 [[a]]$
being the obstruction to the lift.  More precisely, the
inequality $d_1 [[a]] \not= 0$ in $E_1$ means $[db]
\not= d_0 [c]$ in $E_0$. Thus, $db$ cannot be written as a
$\gamma$-variation, even up to the exterior derivative of a
$\gamma$- closed term (ambiguity in the definition of $b$).

Analogous to the decomposition (\ref{iso1}), one has
\bb E_1 \simeq Ker \, d_1 \oplus F_1
\ee where $F_1$ is a subspace of $E_1$ supplementary to
$Ker \, d_1$. The elements in $Im \, d_1$ are trivial in
$H(\gamma \vert d)$ and thus of no interest from the point
of view of the Wess-Zumino consistency condition.

To investigate the (non-trivial) $\gamma$-cocycle that can
be lifted at least three times, one defines
\bb E_2 = H(d_1, E_1)
\ee and the differential
$d_2$ through
\bb d_2 : \; E_2 \rightarrow E_2 , \; d_2 [[[a]]] = [[[dc]]]
\ee where the triple brackets denote the classes in $E_2$
and where
$c$ is defined through the successive lifts
$da + \gamma b = 0$, $db + \gamma c = 0$ (which exist since
$d_1 [[a]] =0$).  It is a straightforward exercise to
verify that $d_2$ is well-defined in $E_2$, i.e., that the
ambiguities in $b$ and $c$ play no r\^ole in $E_2$. 
Furthermore, a $\gamma$-cocycle $a$ such that
$d_0 [a] = 0$ (so that $[[a]] \in E_1$ is well-defined) and
$d_1 [[a]] = 0$ (so that $[[[a]]] \in E_2$ is well-defined)
can be lifted a third time if and only if $d_2 [[[a]]] =0$.
Indeed, the relation $d_2 [[[a]]] =0$ is equivalent to
$[[[dc]]] = 0$, i.e. $dc = \gamma u + d v + d w$, with
$\gamma v = 0$ (this is the $d_0$-term) and $\gamma w + dt
= 0$,
$\gamma t = 0$ (this is the $d_1$-term).  Thus, by
redefining $b$ as
$b - t$ and $c$ as $c -v - w$, one gets, $dc_{Redefined} =
\gamma u$.

\subsection{Lifts of elements of $H(\gamma)$ - General
theory}

One can  proceed in the same way for the next lifts. One
finds in that manner a sequence of spaces $E_r$ and
differentials $d_r$ with the properties

\begin{enumerate}
\item $E_{r} = H(E_{r-1}, d_{r-1})$.
\item There exists an antiderivation $d_r : E_r \rightarrow
E_r$ defined by $d_r [[\dots [ X ] \dots ]] = [[\dots
[db]\dots ]]$ for
$[[\dots [X] \dots ]] \in E_r$ where $[[\dots [db]
\dots ]]$ is the class of the $\gamma$-cocycle $db$ in
$E_r$ and where $b$ is defined through
$dX+ \gamma c_1 = 0$, $dc_1 + \gamma c_2 = 0$, ...,
$d c_{r-1} + \gamma b = 0$. Similarly, $[[\dots [X] \dots
]]$ denotes the class of the
$\gamma$-cocycle $X$ in $E_r$ (assumed  to fulfill the
successive conditions $d_0 [X] = 0$, $d_1 [[X]] =0$ etc ...
so as to define an element of $E_r$).
\item $d_r^2 = 0$. 
\item A $\gamma$-cocycle $X$ can be lifted $r$ times if and
only if
$d_0 [X] = 0$,
$d_1 [[X]]=0$,
$d_2 [[[X]]]=0$,
 ..., $d_{r-1} [[\dots[X]\dots]]=0$.  If $d_r [[\dots[ X
]\dots]]
\not=0$, the $\gamma$-cocycle $X$ cannot be lifted
$(r+1)$ times and is not exact in $H(\gamma \vert d)$.
\item A necessary and sufficient condition for an element
$Y$ in
$H(\gamma)$ to be exact in
$H(\gamma \vert d)$ is that there exists a $k$ such that
$d_i [\dots[Y]\dots]=0$, ($i=1,2, \dots, k-1$) {\em and}
$[\dots[Y]\dots] = d_k [\dots[Z]\dots]$. This implies in
particular
$d_j [\dots[ Y ]\dots] = 0$ for all $j$'s.
\item Conversely, if a $\gamma$-cocycle $Y$ fulfills
$d_i [\dots[Y]\dots]=0$ for $i=0, 1, ..., k-1$ and
$d_k [\dots[Y]\dots] \not= 0$, then, it is not exact in
$H(\gamma \vert d)$.  The condition is not necessary,
however, because there are elements of $H(\gamma)$ that are
non trivial in
$H(\gamma \vert d)$ but which are annihilated by all
$d_i$'s.  This is due to the fact that there are no
exterior form of degree higher than the spacetime
dimension.  We shall come back to this point below.
\end{enumerate}

The meaning of the integer $k$ for which $ Y = d_k Z$ in
item 5 (with
$Y \not= d_i$(something) for $i<k$) is as follows (we shall
drop the multiple brackets when no confusion can arise). 
If the
$\gamma$-cocycle $a$ is  in $Im \, d_0$, then $a = db +
\gamma c$, where $b$ is also a $\gamma$-cocycle. If $a$ is
a non-zero element of
$E_1$ in the image of $d_1$, then again $a = db + \gamma
c$, but $b$ is now {\em not} a cocycle of $\gamma$ since
$a$ would then be in $Im
\, d_0$ and thus zero in $E_1$. Instead, one has $\gamma b
+ d \beta = 0$ where
$\beta$ {\em is} a cocycle of $\gamma$ ($\gamma \beta = 0$)
which is not trivial in $H(\gamma
\vert d)$.  More generally, $k$ characterizes the length of
the descent below
$b$ in $a = db + \gamma c$, $\gamma b + d \beta = 0$ etc.

The proof of items 1 to 4 proceeds recursively.  Assume
that the differential algebras $(E_i,d_i)$ have been
constructed up to order
$r-1$, with the properties 2 through 4.  Then, one defines
the next space $E_r$ as in 1.  Let $x$ be an element of
$E_r$, and let
$X$ be one of the $\gamma$-cocycles such that the class
$[[\dots [ X]
\dots ]]$ in $E_r$ is precisely $x$. Since $X$ can be
lifted $r$ times, one has a sequence  $dX + \gamma c_1 =0$,
..., $dc_{r-1} +
\gamma b =0$. The ambiguity in $X$ is $X \rightarrow X +
\gamma a +du_0 + du_1 + \cdots + du_{r-1}$, where $u_0$ is
a $\gamma$-cocycle (this is the $d_0$-exact term), $u_1$ is
the first lift of a
$\gamma$-cocycle (this is the $d_1$-exact term) etc. 
Setting $u = u_0 + u_1 + \cdots u_{r-1}$, one sees that the
ambiguity in $X$ is of the form $X \rightarrow X +
\gamma a + du$.  On the other hand, the ambiguity in the
successive lifts takes the form $c_1 \rightarrow c_1 +
m_1$, where $m_1$ is a
$\gamma$-cocycle that can be lifted $r-1$ times, $c_2
\rightarrow c_2 + n_1 + m_2$, where $n_1$ descends on $m_1$
and $m_2$ is a
$\gamma$-cocycle that can be lifted
$r-2$ times, ..., and finally $b \rightarrow b + a_1 + a_2
+ \cdots + a_{r-1} + a_r$, where $a_1$ descends $(r-1)$
times, on $m_1$, $a_2$ descends $(r-2)$ times, on $m_2$,
etc, and $a_r$ is a
$\gamma$-cocycle.

The element $X_r \equiv db$ is clearly a cocycle of
$\gamma$, which is annihilated by $d_0$ and the successive
derivations $d_k$ because
$dX_r=0$ exactly and not just up to $\gamma$-exact terms. 
The ambiguity in the successive lifts of $X$ plays no
r\^ole in the class of $X_r$ in $E_r$, since it can
(suggestively) be written $db
\rightarrow db + d_{r-1} m_1 + d_{r-2} m_2 + \cdots + d_1
m_{r-1} + d_0 a_r$.  Thus, the map
$d_r$ is well-defined as a map from $E_r$ to $E_r$.  It is
clearly nilpotent since $dX_r=0$.  It is also a derivation,
because one may rewrite the lift equations for $X$ as
$\tilde{\gamma}(X+ c_1 + c_2 +
\dots + b)=d_r X$ where
\bb
\tilde{\gamma} = \gamma + d.
\ee Let $Y$ be another element of
$E_r$ and $e_1$,  $e_2$, ...$\beta$ its successive lifts.
Then,
$\tilde{\gamma} (Y+ e_1 + e_2 + \dots + \beta) = d_r Y$.
Because
$\tilde{\gamma}$ is a derivation, one has
$\tilde{\gamma}[(X+c_1 + \dots + b)(Y+e_1+\dots
+\beta)]=(d_rX)Y +(-1)^{\epsilon_X} X d_rY + $ forms of
higher form-degree, which implies
$d_r(XY)=(d_rX)Y + (-1)^{\epsilon_X} X d_r Y$: $d_r$ is
also an odd derivation and thus a differential.  This
establishes properties 2 and 3.

To prove property 4, one observes that $X$ can be lifted
once more if and only if one may choose its successive
lifts so that $db$ is
$\gamma$-exact. This is equivalent to stating that $d_r X$
is zero in
$E_r$. Properties 5 and 6 are rather obvious: if $a$ is a
$\gamma$-cocycle which is exact in $H(\gamma \vert d)$, $a=
db+
\gamma c$, then $a=d_k m$ where $k$ is the length of the
descent associated with $\gamma b + dn =0$, which has
bottom $m$.

As shown in \cite{DVTV0,DVTV}, the above construction may be
elegantly captured in an exact couple \cite{Massey}.  The
detailed analysis of this exact couple and the proof of the
above results using directly the powerful tools offered by
this couple may be found in
\cite{DVTV,DVTV1,MT}.

One has, for each $r$, the vector space isomorphisms
\bb E_r \simeq Ker \, d_r \oplus F_r \simeq Im \, d_r
\oplus E_{r+1}
\oplus F_r
\label{iso2}
\ee where $F_r$ is a subspace supplementary to $Ker \, d_r$
in $E_r$. Thus
\bb E_0 \simeq \oplus_{k=0}^{k=r-1} F_k
\oplus_{k=0}^{k=r-1} Im \, d_k
\oplus E_r
\ee Because there is no form of degree higher than the
spacetime dimension,
$d_n =0$ ($d_n a$ has form-degree equal to
$FormDeg(a)+n+1$). Therefore, $E_{n} = E_{n+1} = E_{n+2} =
\dots$. This implies
\bb E_0 \simeq \oplus_{k=0}^{k=n-1} F_k
\oplus_{k=0}^{k=n-1} Im \, d_k
\oplus E_n.
\label{isofinal}
\ee The elements in any one of the $F_k$'s are non trivial
bottoms of the descent which can be lifted exactly $k$
times.  All the elements above them in the descent are also
non trivial solutions of the Wess-Zumino consistency
condition. The elements in $Im \, d_k$ are bottoms which
are trivial in $H(\gamma \vert d)$ and which define
therefore  trivial solutions of the Wess-Zumino consistency
condition. Finally, the elements in $E_n$ are bottoms that
can be lifted all the way up to form degree $n$.  These are
non trivial in $H(\gamma \vert d)$, since they are not
equal to $d_i m$ for some $i$ and $m$. The difference
between the elements in $\oplus F_k$ and those in
$E_n$ is that the former ones cannot be lifted all the way
up to form-degree $n$: one meets an obstruction before,
which is $d_{k} a$ (if $a \in F_k$). By contrast, the
elements in $E_n$ can be lifted all the way up to form
degree
$n$.  This somewhat unpleasant distinction between
$\gamma$-cocycles that are non-trivial in $H(\gamma \vert
d)$ will be removed below, where we shall assign an
obstruction to the elements of $E_n$ in some appropriate
higher dimensional space.

In order to solve the Wess-Zumino consistency condition,
our task now is to determine explicitly the spaces $E_r$
and $F_r$.

\section{Covariant Poincar\'e lemma -- Small algebra}
\setcounter{equation}{0}
\setcounter{theorem}{0}

To that end, we first work out the cohomology of $d_0$ in
$E_0
\equiv H(\gamma)$.  Let $u$ be a $\gamma$-cocycle.  Without
loss of generality, we may assume that $u$  takes the form
\bb u = \sum P_I \omega^I
\ee where the $\omega^I$ form a basis of the algebra
generated by the last, non trivial (if necessary, improved)
ghosts of ghosts in the cohomology (as well as $trC^5$,
$trC^7$ etc for the fourth Chapline-Manton model), and
where the $P_I$'s are polynomials in the (improved) field
strength components and their (covariant) derivatives, with
coefficients that involve $dx^\mu$.  The $P_I$'s are
called  ``gauge-invariant polynomials". A direct
calculation using the fact that the $d$-variation of the
last ghosts (and $trC^5$,
$trC^7$ etc) is $\gamma$-exact yields $du = \sum (dP_I)
\omega^I +
\gamma v'$.  This is
$\gamma$-exact if and only if $dP_I =0$.

Now, if $P_I = dR_I$ where $R_I$ is also a gauge invariant
polynomial, then $u$ is $d$-exact modulo $\gamma$, $u=da +
\gamma b$, with
$\gamma a = 0$.  Conversely, if $u =da + \gamma b$ with
$\gamma a = 0$, then
$P_I$ is $d$-exact in the space of invariant polynomials. 
Thus, the class of $u$ (in $E_0$) is a non trivial cocycle
of $d_0$ if and only if $P_I$ is a non trivial cocycle of
the {\em invariant} cohomology of $d$.  We give below the
relevant ``covariant Poincar\'e lemma" for each of the
models of this paper.

Since we are interested in lifts of $\gamma$-cocycles from
form-degree $k$ to form-degree $k+1$, we shall investigate
the
$d$-invariant cohomology only in form-degree strictly
smaller than the spacetime dimension $n$.  This will be
assumed throughout the remainder of this section. [In
form-degree $n$, there is clearly further invariant
cohomology since any invariant $n$-form is annihilated by
$d$, even when it cannot be written as the $d$ of an
invariant form].

\subsection{Free case} The calculation of the invariant
cohomology of
$d$ is given in the appendix A of \cite{HKS1}, so we just
recall the result.
\begin{theorem} The invariant cohomology of $d$ for the
free theory is given by the polynomials in the exterior
forms $H^a$.
\end{theorem} This is a direct generalization of the result
established for $1$-forms in \cite{BDK1,DVHTV}. Note that
the cohomology contains in particular the constants and the
constant forms.  These latter can be eliminated by imposing
Lorentz invariance.

The theorem implies, according to the general analysis of
the descent equation given above, that the only bottoms $u$
($\gamma u = 0$) that can be lifted at least once can be
expressed in terms of exterior products of the curvature
forms $H^a$ and the last ghosts of ghosts (up to trivial
redefinitions). Out of the infinitely many generators of
$H(\gamma)$, only $H^a$ and
$B^{a(0,p_a)}$ survive in $E_1$.

Because the objects that survive the first step in the lift
can be expressed in terms of forms, it is convenient to
introduce the so-called ``small algebra"
${\cal A}$ generated in the exterior product by the
exterior forms
$B^{a(k,p_a-k)}$ and $dB^{a(k,p_a-k)}$ ($k=0, ..., p_a$).
This algebra is stable under
$\gamma$ and $d$.  If one denotes by $E^{small}_0$ the
cohomology of
$\gamma$ in the small algebra, one finds
\bb E_0^{small} \equiv H(\gamma, {\cal A}) \simeq {\cal B}
\ee where ${\cal B}$ is the subalgebra of ${\cal A}$
generated by the curvatures $H^a$ and the last ghosts of
ghosts $B^{a(0,p_a)}$.

One defines $E_1^{small}$ as $H(d_0^{small}, E_0^{small})$,
where
$d_0^{small}$ is the restriction of $d_0$ to $E_0^{small}$.
Because
$d H^a = 0$ and $dB^{a(0,p_a)} = \gamma$(something), the
restriction
$d_0^{small}$ identically vanishes. Thus
\bb E_1^{small}
\simeq E_0^{small} \simeq {\cal B}.
\ee

What is the relationship between $E_1^{small}$ and $E_1$? 
These two spaces are in fact isomorphic,
\bb E_1 \simeq E_1^{small}.
\label{E1E1small}
\ee Indeed, let $q$ be the map from $E_1^{small}$ to $E_1$
that assigns to a cohomological class in $E_1^{small}$ its
cohomological class in $E_1$ ($a \in  E_1^{small} \simeq
{\cal B}$ fulfills $\gamma a = 0$ and $d_0 a =0$ and thus
defines of course an element of $E_1$). It follows from the
above theorem that the map $q$ is surjective since any
class in $E_1$ possesses a representative in the small
algebra.  The map $q$ is also injective because there is no
non trivial class in $E_1^{small}$ that becomes trivial in
$E_1$. If the small-algebra $\gamma$-cocycle
$r = \sum P_I \omega^I$ with $P_I, \omega^I \in {\cal B}$
can be written as $r = du + \gamma t$ where $u$ and $v$ are
in the big algebra and $u$ is a $\gamma$-cocycle, then
$r$ is actually zero.  This can be easily seen by setting
the derivatives of the ghosts and of the field strength
components equal to zero.

The differentials $d_1^{small}$ and $d_1$ are mapped on
each other in this isomorphism,
$d_1 q = q d_1^{small}$. It then follows that the next
cohomological  spaces $E_k$ and $E_k^{small}$ are also
equal,
$E_2 \equiv H(d_1, E_1) \simeq E_2^{small}
\equiv H(d_1^{small}, E_1^{small})$, $E_3 \simeq
E_3^{small}$ etc (but of course, $E_0 \not= E_0^{small}$).
{\em There is thus no loss of generality in investigating
in the small algebra the solutions of the Wess-Zumino
consistency condition that descend non trivially}.

\subsection{Chapline-Manton models}

The small algebra ${\cal A}$ is also relevant to the
Chapline-Manton models because the invariant cohomology can
be computed in it without loosing any cohomological class. 
Indeed, one has

\begin{theorem} Let $P$ be a gauge invariant polynomial. If
$P$ is closed, then $P$ is the sum of a closed, gauge
invariant polynomial belonging to the small algebra and of
the exterior derivative of an invariant polynomial,
\bb dP = 0  \Leftrightarrow P=Q+dR, \; Q \in {\cal A}, \;
dQ=0
\label{invariantpol}
\ee (with $P$, $Q$ and $R$ all gauge-invariant). 
Furthermore, if
$Q$ is $d$-exact in the algebra of gauge-invariant
polynomials, $Q= dS$ with $S$ gauge-invariant, one may
assume that $S$ is in the small algebra (and
gauge-invariant). Therefore, the invariant cohomology of
$d$ is to be found in ${\cal A}$.
\end{theorem}

Note that while the conditions $Q \in {\cal A}$ and $Q=dS$
(with $S$ gauge-invariant) imply
$Q=0$ in the free case, this is no longer true here.

We shall prove the theorem for the specific case of the
second model.  The proof proceeds in the same way for the
other models. Introduce a grading $N$ that counts the
number of derivatives of the
$B$ field.  According to this grading $P$ and $d$ split as
\bb P = P_k + P_{k-1} + \cdots + P_1 + P_0, \; \; d = D_1 +
D_0,
\ee with
\bb N(P_i) = i, \; \; N(D_i)=i.
\ee The differential $D_1$ takes derivatives only of the
$B$-field, the differential $D_0$ takes derivatives only of
the $A$-field. Because $P$ is gauge-invariant, the
$B$-field enters $P$ only through the components of $dB$
and their derivatives.  Furthermore, even though the
$P_i$'s with $i<k$ may involve the components
$A_\mu$'s and their symmetrized derivatives, $P_k$ depends
on $A$ only through the $F_{\mu \nu}$ and their derivatives.

The equation $dP = 0$ yields $D_1 P_k =0$ at the highest
value of the
$N$-degree. According to the results for the free case,
this implies
$P_k$ = $D_1 R_{k-1} + m_k$ where $R_{k-1}$ is a polynomial
in the components of $dB$ and their derivatives, while
$m_k$ is a polynomial in the form $dB$, both with
coefficients in the components of $F$ and their derivatives
(which fulfill
$D_1 F_{\mu \nu} =0$).  Covariantize $R_{k-1}$ and $m_k$ by
completing
$dB$ into $H$.  This only introduces terms of lower
$N$-degree.  We denote the covariant objects by $r$ and
$m$, respectively. One has
$P_k = (dr + m)_k$ and $P = dR_{k-1} + m_k + \; more$,
where ``$more$" is an invariant polynomial of maximum
$N$-degree strictly smaller than $k$. The invariant
polynomial $m$ - which exists only if $k=1$ or $0$ since
$H^2 = 0$ - is of order $k$ in the exterior form $H$.  It
must be closed by itself since there can be no compensation
between $D_0 m$ and $D_1(more)$, which is necessarily of
lower degree in the components of $H$ and their
derivatives.  It follows from $D_0 m = 0$ that $m =
\mu(F,H) + ds$, where $\mu$ is a polynomial in the forms
$F$ and $H$ and where $s$ is an invariant polynomial (use
again the results for the free case and
$H dp = -d(Hp)+more$).  Thus one can get rid of $P_k$ by
adding to $P$ terms of the form (\ref{invariantpol}) of the
theorem.   By repeating the argument at the successive
lower degrees, one reaches the desired conclusion.

To prove the second part of the theorem, one first observes
that if
$dQ=0$, then $Q(F,H)$ does not involve in fact $H$,
$Q=Q(F)$ (see subsection 5.2.2 below).  Assume then that
$Q=dU$, where $U$ is a gauge-invariant polynomial,
$U=U([H],[F])$. By expanding $U$ according to the
$N$-degree, $U=U_0 + U_1 +... + U_l$, one finds at higher
order $D_1 U_l = 0$, which implies as above
$U_l = D_1 R_{l-1} + m_l$ where $m_l$ is a polynomial in
the form
$dB$.  One can remove $D_1 R_{l-1}$ from $U_l$ by
substracting
$dR_{l-1}$ from $U$, which does not modify $Q$.  Thus, only
$m_l$, which is present for
$l=1$ or $l=0$, is relevant.  By repeating the argument,
one finally arrives at
\bb U = H a([F]) + b([F])
\ee The condition $Q = dU$ implies $da=0$ and thus $a = d
\nu([F]) +
\rho(F)$ where $\rho(F)$ is a polynomial in the form $F$.
The term $H d \nu([F])$ is irrelevant since it can be
absorbed into $b([F])$ up to a $d$-exact term.  Thus, $U =
H \rho(F) + b'([F])$.  The condition
$Q(F) = dU$ reads now $Q(F) = k(F) + db'([F])$ where $k(F)$
is a polynomial in $F$ and implies $db' = 0$ because
$db'([F])$ necessarily involves one derivative of
$F_{\lambda \mu}$ if it is not zero.  But then, again, one
can drop
$b'$ from $U$, which proves the second assertion.

It follows from this theorem that there is no restriction in
investigating the invariant $d$-cohomology in the small
algebra.  Elements of $H(\gamma)$ that can be lifted at
least once necessarily belong to ${\cal A}$ up to trivial
terms. There is no restriction in the investigation of the
next lifts either because again $E_1^{small} = E_1$. If a
$\gamma$-cocycle $a \in  {\cal A}$ can be written as $a =
du + \gamma v$ where
$u$ and $v$ are in the big algebra and $\gamma u = 0$, then
one may find $u'$ and $v'$ in ${\cal A}$ such that
$a=du'+\gamma v'$ (with
$\gamma u'=0$).  This follows from the second part of the
theorem. Obstructions to lifts within ${\cal A}$ are not
removed by going to the big algebra.

\subsubsection{Model 1} For the first Chapline-Manton model
discussed above, the invariant cohomology of $d$ is
trivial. Indeed, in the algebra generated by $F$ and $H$,
the differential $d$ takes the contractible form $dF = H$,
$dH= 0$. Thus
\bb E_1 \equiv H(d_0, E_0^{small}) = 0
\ee where $E_0^{small}$ is the algebra generated by $F$ and
$H$.

\subsubsection{Model 2} In the algebra generated by the
gauge-invariant curvatures,
$d$ takes the form
\bb dF = 0, \; dH = F^{r+1}.
\ee Since $H^2=0$, any element in this algebra is of the
form
\bb a = \alpha(F) + \beta(F) H
\ee where $\alpha(F)$ and $\beta(F)$ are polynomials in F.
The condition that $a$ is closed implies $\beta(F)F^{k+1}
=0$, which forces $\beta (F)$ to vanish.  Furthermore $a
\equiv \alpha(F)$ is exact if it is in the ideal generated
by $F^{r+1}$. Thus, we have the theorem:
\begin{theorem} The invariant cohomology of $d$ for the
Chapline-Manton mo\-del 2 is the quotient of the algebra
generated by the $F$'s by the ideal generated by $F^{r+1}$.
\end{theorem}

\subsubsection{Model 3} For the third model, $d$ is given by
(\ref{Bianchi5}). By redefining the curvature $G$ as
\bb G_M = G- \frac{F^2}{2}
\ee this can be brought to the form
\bb dF = -H, \; dH=0, \; dG_M = 0
\ee from which it follows that:
\begin{theorem} For the third model, the invariant
cohomology of $d$ is given by the polynomials in the
variable $G_M=G-F^2/2$.
\end{theorem}

\subsubsection{Model 4}

The invariant polynomials in the small algebra are the
polynomials in the gauge-invariant curvature $H$ of the
$2$-form and in the ``fundamental" invariants
$tr F^2$, $tr F^3$, ... $tr F^N$ for $SU(N)$ (this is a
basis for the
$SU(N)$ symmetric polynomials). These polynomials are
closed, except
$H$, which fulfills
$dH = tr F^2$.  Hence,
$H$ and  $tr F^2$ do not appear in the cohomology.

\begin{theorem} For the fourth model, the invariant
cohomology of $d$ is given by the polynomials in $tr F^3$,
$tr F^4$, ... $tr F^N$.
\end{theorem}

\subsection{Universal algebra ${\cal U}$} The small algebra
${\cal A}$ involves only exterior forms, exterior products
and exterior derivatives.   It does ``remember" the
spacetime dimension since its generators are not free: any
product of generators whth form-degree exceeding the
spacetime dimension vanishes.

It is useful to drop this relation and to work in the
algebra freely generated by the potentials, the last ghosts
of ghosts and their exterior derivatives with the sole
condition that these commute or anti-commute (graded
commutative algebra) but without imposing any restriction
on the maximally allowed form degree
\cite{DVTV,Bonora}. This algebra is called the universal
algebra and denoted by ${\cal U}$.  In this algebra, the
cohomology of $d$ is trivial in all form-degrees and the
previous theorems on the invariant cohomology of $d$ are
also valid in form-degree
$\geq n$. Furthermore, one can sharpen the condition for a
cocycle in
$H(\gamma)$ to be non trivial in $H(\gamma \vert d)$.

\begin{theorem} A necessary and sufficient condition for $X
\in H(\gamma)$ to be non-trivial in $H(\gamma \vert d)$ is
that there exists
$r$ such that $d_rX \not=0$.  That is, the lift of $X$ must
be obstructed at some stage. (For the equation $d_rX
\not=0$ to make sense, $d_i X$ must vanish for $i<r$. 
Also, we denote again $X \in E_0$ and its representative in
$E_r$ by the same letter).
\end{theorem}

Proof: The decomposition of $E_n$ is now non-trivial since
$da$ does not necessarily vanish even when $a$ is a
$n$-form. Thus,
$d_n$ is not necessarily zero and the procedure of lifting
can be pursued above form-degree $n$. Suppose that one does
not encounter an obstruction in the lifting of $X$. That
is, one can go all the way up to ghost number zero, the
last two equations being $dc_k + \gamma b = 0$ (with $b$ of
ghost number zero) and
$db = 0$ (so $b$ lifts to zero).  Then, one can write $b =
d m$ since the cohomology of $d$ is trivial in any
form-degree in the universal algebra ${\cal U}$ (except for
the constants, which cannot arise here since $b$ involves
the fields). The triviality of the top-form $b$ implies the
triviality in $H(\gamma \vert d)$ of all the elements below
it.  Thus, a necessary condition for the bottom to be non
trivial in $H(\gamma \vert d)$ is that one meets an
obstruction in the lift at some stage. The condition is
also clearly sufficient.

One can summarize our results as follows

\begin{theorem} (Generalized ``transgression" lemma) Let $X
\in E_0$ be a non-trivial element of
$H(\gamma \vert d)$.  Then there exists an integer $r$ such
that $d_i X = 0$, $i<r$ and $d_rX = Y \not=0$.  The element
$Y$ is defined through the chain $dX+ \gamma c_1 = 0$, ...,
$dc_{r-1} + \gamma c_r=0, dc_r + \gamma c_{r+1}= Y$, where
the elements $c_i \in {\cal U}$ ($i=1, r+1$) are chosen so
as to go all the way up to $c_{r+1}$. One has $\gamma Y=0$
and $Y$ should properly be viewed as an element of
$E_r$ (reflecting the ambiguities in the lift). One calls
the obstruction $Y$ to a further lift of $X$ the
(generalized) ``transgression" of $X$.  The element $X$ and
its transgression have opposite statistics.
\end{theorem} This is the direct generalization of the
analysis of
\cite{DVTV} to the case of $p$-forms.  ``Primitive
elements" of $E_0$ are those that have form-degree zero and
for which the transgression has ghost number zero, i.e.,
they are the elements that can be lifted all the way up to
ghost number zero (``that can be transgressed").  We refer
to \cite{DVTV,Greub,Transgr} for more background
information applicable to the Yang-Mills case.

Because the space $E_n$ and the next ones can be further
decomposed in the universal algebra,
\bb E_n \simeq Im d_{n} \oplus E_{n+1} \oplus F_n, \;
E_{n+1} \simeq etc
\ee where the decomposition for a given
$\gamma$-cocycle ultimately ends at form-degree equal to
the ghost number, one has
\bb E_0 \simeq \oplus_{k=0}^\infty F_k \oplus_{k=0}^\infty
Im d_{k}.
\ee

\section{Results}
\label{results}
\setcounter{equation}{0}
\setcounter{theorem}{0}

We can now compute the spaces $E_k$.

\subsection{Free case}

Let $0<p_1<p_2< \dots <p_M$ be the form degrees of the gauge
potentials $B^a$.  We denote by $B_1^{a_1}$ the forms of
degrees
$p_1$, $B_2^{a_2}$ the forms of degree
$p_2$ etc.

The first non-vanishing differential (in $E_0^{small}$) is
$d_{p_1} $ so that $E_0^{small} = E_1 = E_2 = ... =
E_{p_1}$. Any bottom in
$E_0^{small}$ can be lifted at least $p_1$ times.  In
$E_1$, the differential $d_{p_1}$ acts as follows
\bb d_{p_1} B_1^{a_1(0, p_1)} = H_1^{a_1}, \; d_{p_1}
H_1^{a_1} = 0
\ee in the sector of the forms of degree $p_1$ and
\bb d_{p_1} B_k^{a_k(0, p_k)} = 0, \; d_{p_1} H_k^{a_k} =
0, \; k >1
\ee in the other sectors.  The form of the differential
$d_{p_1}$ makes explicit the contractible part of
$(E_{p_1}, d_{p_1})$. The variables $B_1^{a_1(0, p_1)}$ and
$H_1^{a_1}$  are removed from the cohomology, so that
$E_{p_1+1}$ is isomorphic to the algebra generated by the
curvatures $H_k^{a_k}$ of form-degree
$>p_1 +1$ and the last ghosts of ghosts of ghost number
$>p_1$.

A subspace $F_{p_1}$ complementary to $Ker \, d_{p_1}$ is
easily constructed.  In fact, a monomial in
$B_1^{a_1(0, p_1)}$ and $H_1^{a_1}$ is defined by a tensor
$f_{a_1 \dots a_k b_1 \dots b_m}$ which is symmetric
(respectively antisymmetric) in $a_1, \dots, a_k$ and
antisymmetric (respectively symmetric) in $b_1, \dots, b_m$
if the last ghosts of ghosts are commuting (respectively
anticommuting).  Its irreducible components can be of two
Young-symmetry types, one of which must be zero if the
monomial is to be annihilated by $d_{p_1}$.  The space
$F_{p_1}$ can be taken to be the space generated by the
monomials of this symmetry type (not annihilated by
$d_{p_1}$), tensored by the algebra generated by the
curvatures and last ghosts of ghosts of higher degree.
Together with their successive lifts, the elements in
$F_{p_1}$ provide all the non-trivial solutions of the
Wess-Zumino consistency condition which are involved in
descents whose bottoms can be lifted exactly $p_1$ times.

Similarly, one finds that the next non-vanishing
differential is
$d_{p_2}$.  The generators $B_2^{a_2(0, p_2)}$ and
$H_2^{a_2}$ drop from the cohomology of $d_{p_2}$ while
those of higher degree remain.  A space $F_{p_2}$ can be
constructed along exactly the same lines as the space
$F_{p_1}$ above and characterizes the solutions of the
Wess-Zumino consistency condition involved in descents
whose bottoms can be lifted exactly $p_2$ times.

More generally, the non-vanishing differentials are
$d_{p_k}$. They are defined (in $E_{p_k}$, which is
isomorphic to the algebra generated by the curvatures of
form-degree $>p_{k-1}+1$ and the last ghosts of ghosts of
ghost number $>p_{k-1}$) through
\bb d_{p_k}B_k^{a_k(0, p_k)}=H_k^{a_k}, \; d_{p_k}
H_k^{a_k} = 0
\ee and
\bb d_{p_k} B_j^{a_j(0, p_j)} = 0, \; d_{p_k}H_j^{a_j} = 0,
\; j>k.
\ee The generators $B_k^{a_k(0, p_k)}$ and $H_k^{a_k}$
disappear in cohomology.  The subspace $F_{p_k}$ is again
easily constructed along the previous lines. Together with
their successive lifts, the elements in $F_{p_k}$ provide
all the non-trivial solutions of the Wess-Zumino
consistency condition which are involved in descents whose
bottoms can be lifted exactly $p_k$ times.

The discussion can be illustrated in the case of the simple
model with one free $1$-form $A^{(1,0)}$ and one free
$2$-form $B^{(2,0)}$ considered in subsection 4.2.  The
space $E_0^{small}$ is isomorphic to the space of
polynomials in the curvature-forms $F$, $H$ and the last
ghosts of ghosts $A^{(0,1)}$, $B^{(0,2)}$.  The differential
$d_0^{small}$ vanishes so $E_1 = E_0^{small}$.  One finds
next $d_1 A^{(0,1)} = F$,
$d_1 F = 0$, $d_1 B^{(0,2)} = 0$ and $d_1 H=0$.  The space
$E_2$ is isomorphic to the space of polynomials in
$B^{(0,2)}$ and $H$.  One may take for
$F_1$ the space of polynomials linear in $A^{(0,1)}$. 
These can be lifted exactly once, their lifts being linear
in
$A^{(1,0)}$ and $A^{(0,1)}$,
\bb a \in F_1 \Leftrightarrow a = A^{(0,1)} \sum
(B^{(0,2)})^l F^k H^m \; \; (m=0 \hbox{ or } 1)
\ee Then, one gets
\bb da + \gamma b = 0
\ee with
\bb b = \sum \big( A^{(1,0)} (B^{(0,2)})^l F^k H^m + l
A^{(0,1)} B^{(1,1)} (B^{(0,2)})^{l-1} F^k H^m  \big).
\ee They cannot be lifted a second time since the
obstruction $d_1 a =\sum (B^{(0,2)})^l$ $F^{k+1} H^m$ does
not vanish.  The above $a$'s and
$b$'s are the most general solutions of the Wess-Zumino
consistency condition involved in descents of length $1$.

The differential $d_2$ in $E_2$ is given by
$d_2 B^{(0,2)} = H$, $d_2 H = 0$.  Because $H^2=0$, one may
take for
$F_2$ the space of polynomials in $B^{(0,2)}$ only.  For
those, the descent reads,
\begin{eqnarray}
\alpha = (B^{(0,2)})^l &,& \gamma \alpha = 0
\nonumber \\
\beta = l B^{(1,1)} (B^{(0,2)})^{l-1} &,& d \alpha +\gamma
\beta = 0
\nonumber \\
\lambda = l B^{(2,0)} (B^{(0,2)})^{l-1} + \frac{l(l-1)}{2}
(B^{(0,2)})^{l-2} (B^{(1,1)})^2 &,& d \beta+\gamma \lambda 
= 0
\end{eqnarray} The elements of the form $\alpha$, $\beta$
or $\lambda$ are the most general solutions of the
Wess-Zumino consistency condition involved in descents of
length $2$.  With the solutions involved in descents of
length $1$ and those that do not descend (i.e., which are
strictly annihilated by $\gamma$), they exhaust all the
(antifield-independent) solutions of the Wess-Zumino
consistency condition.

A straightforward consequence of our discussion is the
following theorem, which will prove useful in \cite{HK}.
\begin{theorem}
\label{triviality} Let $\omega$ be a $\gamma$-cocycle that
takes the form
\bb
\omega = \alpha(H_s^{a_s} \; B_s^{a_s,(0, p_s)})
\beta(B_k^{a_k(0, p_k)}, H_k^{a_k}), \; k>s
\ee where $\alpha$ vanishes if $H_s^{a_s}$ and $B_s^{a_s(0,
p_s)}$ are set equal to zero (no constant term) and fulfills
\bb d_{p_s} \alpha = 0
\ee (the first potential obstruction in the lift of
$\omega$ is absent). Then, $\omega$ is trivial in $H(\gamma
\vert d)$.
\end{theorem} The proof is direct, one has $\alpha =
d_{p_s} \mu$ for some
$\mu(H_s^{a_s}, B_s^{a_s(0, p_s)})$ since $d_{p_s}$ is
acyclic in the space of the $\alpha(H_s^{a_s} B_s^{a_s,(0,
p_s)})$  with no constant term. Thus $\omega$ is
$d_{p_s}$-exact, $\omega= d_{p_s} (\mu \beta)$: $\omega$ is
the first obstruction to the further lift of $\mu \beta$
and as such, is trivial.

The theorem applies in particular when $\alpha$ is an
arbitrary polynomial of degree $>0$ in the curvatures
$H_s^{a_s}$.

\subsection{Chapline-Manton models}

\subsubsection{Model 1} The analysis is trivial in this
case since there is no non trivial descent.  All solutions
of the Wess-Zumino consistency condition can be taken to be
strictly annihilated by
$\gamma$, i.e., can be taken to be in $E_0$ ($E_1 = 0$). 
They are thus completely described by Theorem {\bf
\ref{gammaCohoCM1}} (from which one must remove the
$d$-exact terms $d\alpha([F])$).

\subsubsection{Model 2} The second model is more
interesting.  The algebra $E_0^{small}$ has generators $H$,
$F$, $A^{(0,1)}$ and
$B^{(0,2r)}$. One may take for $F^{small}_0$ the space of
elements of the form $H Q(F,A^{(0,1)}, B^{(0,2r)})$ for
which
$d_0 (H Q(F,A^{(0,1)}, B^{(0,2r)})) = F^{r+1} Q \not=0$. 
These
$\gamma$-cocycle do not lift at all.  The space $E_1$ is
isomorphic to the algebra generated by $F$, $A^{(0,1)}$ and
$B^{(0,2r)}$, with the relation $F^{r+1} = 0$.  Thus, it is
not a free algebra contrary to the situation encountered in
the free case.  The differential $d_1$ is non trivial and
given by
\bb d_1 A^{(0,1)} = F, \; d_1 F = 0, \; d_1 B^{(0,2r)} = 0
\ee when $r>1$, which we shall assume at first.  Because 
$F$ is subject to the relation $F^{r+1} = 0$, the
cohomological space
$E_2 \equiv H(d_1,E_1)$ is isomorphic to the algebra
generated by
$B^{(0,2r)}$ and $\mu(A,F)$ with
\bb
\mu(A,F) = -A^{(0,1)} F^r
\ee The next differentials $d_2$, $d_3$ ... vanish up to
$d_{2r-1}$. So, $E_2 = E_3 = \dots = E_{2r-1}$.  One has
\bb d_{2r-1} B^{(0,2r)} = \mu(A,F), \; d_{2r-1} \mu(A,F) =
0.
\ee Thus $E_{2r} = 0$.

One can take for $F_1$ the space  of polynomials of the form
$(B^{(0,2r)})^l Q_l(F)$ $A^{(0,1)}$ where $Q_l$ is a
polynomial in
$F$ of degree strictly less than
$r$. Similarly, one may take for  $F_{2r-1}$ the space of
polynomials in
$B^{(0,2r)}$ (with no constant piece).   We leave it to the
reader to write down explicitly the lifts of these
elements.  Note in particular that $\mu(A,F)$ does not
appear in any of the spaces
$F_k$.  This is because it is now trivial.  In the free
case,
$\mu(A,F)$ is an element of $F_1$ and is the bottom of a
non-trivial descent of length two.  The coupling to the
$2$-form makes it disappear from the cohomology. At the
same time, the cocycle
$F^{r+1}$, which is in the invariant cohomology of $d$ in
the free case, has now become $d$-exact in the space of
invariant polynomials.  Also, while $B^{(0,2r)}$ can be
transgressed all the way up to $H$ in the free case, its
lift stops now at ghost number one with $\mu$.

The situation for $r=1$ is similar, the two steps
corresponding to the differentials $d_1$ and $d_{2r-1}$
being combined in one, so that the space $E_2$ vanishes. 
The easiest way to see this is to observe that $H(d_1,E_1)$
(with $d_1 A^{ (0,1)} = F$, $d_1 F = 0$ and $d_1 B^{(0,2)}
= \mu(A,F)$ for $r=1$) is isomorphic to
$H(D,E_0)$ with $DA^{ (0,1)} = F$, $DF = 0$,
$DH = F^{r+1}$, $D B^{(0,2)} = \mu(A,F) + H$.  Indeed, one
may view the generator $H$ as Koszul generator for the
equation
$F^{r+1} = 0$.   The change of variable $H \rightarrow H' =
H + \mu$ brings then $D$ to the manifestly contractible
form.

\subsubsection{Model 3}

The third model is essentially a combination of the first
model in the ($A$, $B$)-sector and of the free model for
the improved $3$-form
$C_M = C - AB- \frac{1}{2} A dA$, with curvature $G_M =
dC_M$ and improved last ghost of ghost
$\tilde{C}^{(0,3)}$ (\ref{improvedGhost}).  Only $d_0$ and
$d_3$ are non trivial.  The details are left as an exercise.

\subsubsection{Model 4}

In the absence of coupling between the $2$-form and the
Yang-Mills field, the non trivial differentials are
\bb d_2 B^{(0,2)} = H, \; d_2 H =0
\label{d2CM4}
\ee ($B^{(0,2)} \equiv \rho$) and
\begin{eqnarray} d_3 trC^3 &=& tr F^2, \; d_3 tr F^2 =0
\label{d3CM4}\\ d_5 trC^5 &=& tr F^3, \; d_5 tr F^3 = 0
\label{d5CM4}\\ &\vdots& \\ d_{2N-1} tr C^{2N-1} &=& tr
F^{N}
\label{dNCM4}
\end{eqnarray} (see \cite{DVTV}).  We have written
explicitly only the action of the non trivial $d_k$'s on
the contractible pairs. The last ghost of ghost $B^{(0,2)}$
is non trivial and can be lifted twice;
$trC^3$ is non trivial and can be lifted three times;
$trC^5$ is non trivial and can be lifted five times; more
generally, $tr C^{2k+1}$ is non trivial and can be lifted
$(2k+1)$ times.

When the coupling is turned on, the variables
$\rho$ and $trC^3$ disappear from the cohomology. It
follows that all the solutions of the Wess-Zumino
consistency condition that were previously above $trC^3$
(or above a polynomial involving $trC^3$) become trivial. 
This is the Green-Schwarz anomaly cancellation mechanism
\cite{GS}. At the same time, the differential $d_0$ becomes
non trivial, as for the previous Chapline-Manton models.
One has
\bb d_0 H = tr F^2, \; d_0 tr F^2 = 0
\ee which shows that $tr F^2$ disappears from the invariant
cohomology, as already pointed out above.  The other
differentials (\ref{d5CM4}) through (\ref{dNCM4}) remain
unchanged. The cohomology
$H(\gamma \vert d)$ is given in \cite{HK2}.

\section{Counterterms and anomalies}
\setcounter{equation}{0}
\setcounter{theorem}{0}

We finally summarize  our results by giving explicitly the
antifield-inde\-pen\-dent counterterms and anomalies, i.e.,
$H^{(n,0)}(\gamma \vert d)$ and $H^{(n,1)}(\gamma \vert
d)$. These can be of two types: (i) the ones that descend
trivially (``type A"); these can be assumed to be strictly
annihilated by $\gamma$ and are described by $H(\gamma)$ up
to trivial terms; and (ii) the ones that lead to a
non-trivial descent (``type B"); these can be assumed to be
in the small algebra modulo solutions of the previous type.
For small ghost number, it turns out to be more convenient
to determine the solutions of ``type B" directly from the
obstructions sitting above them rather than from the
bottom. That this procedure, which works in the universal
algebra, yields all the solutions, is guaranteed by our
general analysis.

\subsection{Counterterms and anomalies of type A}

a) The counterterms that lead to a trivial descent
involve in general the individual components of the
gauge-invariant field strengths and their derivatives and
generically cannot be expressed as exterior products of the
forms $F$ or $H$. They are the gauge-invariant polynomials
introduced above and read explicitly
\bb 
\setlength{\fboxsep}{7pt}
\framebox{$a = a([H^a]) d^n x$}
\setlength{\fboxsep}{3pt}\label{71}
\ee for the free models\footnote{Recall that the ``free
models" encompass in fact all models having the gauge
symmetries of the free theory since the present analysis
depends only on the form of the gauge transformations and
not on the specific Lagrangian.  So the results for the
``free models" cover general  Lagrangians $L([H^a])$ or
models with Chern-Simons interactions.},
\bb 
\setlength{\fboxsep}{7pt}
\framebox{$a = a([F]) d^n x$}
\setlength{\fboxsep}{3pt}\label{72}
\ee for the first Chapline-Manton models,
\bb 
\setlength{\fboxsep}{7pt}
\framebox{$a = a([F],[H]) d^n x$}
\setlength{\fboxsep}{3pt}\label{73}
\ee for the second Chapline-Manton models and
\bb 
\setlength{\fboxsep}{7pt}
\framebox{$a = a([F],[G]) d^n x$}
\setlength{\fboxsep}{3pt}\label{74}
\ee for the third Chapline-Manton models (with the condition
$a \not= db$ in all cases which is equivalent to the
condition that the variational derivatives of $a$ with
respect to the fields do not identically vanish).  

For the
fourth Chapline-Manton model, the counterterms are
invariant functions of $F^a_{\mu
\nu}$ and their covariant derivatives, as well as of $[H]$,
\begin{equation}
\setlength{\fboxsep}{7pt}
\framebox{$a=P_I([F^a],[H])d^n x,$}
\setlength{\fboxsep}{3pt}\label{75}
\end{equation}
where $P_I$ is invariant under the adjoint representation
of $SU(N)$.

We have assumed that the spacetime forms
$dx^\mu$ occur only through the product $dx^0 dx^1 \cdots
dx^{n-1}
\equiv d^n x$ as this is required by Lorentz-invariance.
\vspace{.5cm}

\noindent b) The anomalies that lead to a trivial descent
are sums of terms of the form
$a = P \, C \, d^n x$ where $P$ is a gauge-invariant
polynomial and
$C$ is a last ghost of ghost of ghost number one, which
must be non trivial in $H(\gamma)$.  These anomalies exist
only for a free theory with $1$-forms and in the second
Chapline-Manton models since only in these cases are there
non trivial, last ghosts of ghosts of ghost number one. 
One has explicitly
\bb 
\setlength{\fboxsep}{7pt}
\framebox{$a = P_A([H^a]) B^{A(0,1)},$}
\setlength{\fboxsep}{3pt}
\ee where $A$ runs over the $1$-forms (free models) or
\bb 
\setlength{\fboxsep}{7pt}
\framebox{$a = P([F],[H]) A^{(0,1)},$}
\setlength{\fboxsep}{3pt}
\ee (second CM models). In both cases $a$ will be trivial if
$P=dR$ where $R$ is an invariant polynomial or if
$P_A=P_A(H^a)$ with
$ P_A H^A=0$ in the first case and $a=\mu$ in the second
case.

The existence of such anomalies - which generically cannot
be expressed as exterior products of curvatures and ghosts
- was pointed out in
\cite{DixonRM} for Yang-Mills gauge models with $U(1)$
factors.

\subsection{Counterterms of type B} The solutions that lead
to a non trivial descent can be assumed to be in the small
algebra, i.e., can be expressed in terms of exterior
product of the fields, the ghosts (which are all exterior
forms) and their exterior derivatives (modulo solutions of
type A). If $a$ is a non-trivial solution of the
Wess-Zumino consistency condition with ghost number zero,
then $da
\not= 0$ (in the universal algebra). Since
$a$ has ghost number zero, it  is the top of the descent
and $da$ is the obstruction to a further lift. Because $da$
is a
$\gamma$-cocycle, it is a gauge-invariant polynomial.  It
must, in addition, be $d$-closed but not $d$-exact in the
space of gauge-invariant polynomials since otherwise, $a$
could be redefined to be of type A.  Therefore, $da$ is an
element of the invariant cohomology of $d$ and it will be
easier to determine
$a$ directly from the obstruction
$da$ rather than from the bottom of the descent because one
knows  the invariant cohomology of $d$.

\subsubsection{Free models}

In the free case, any polynomial $P(H)$ in the curvatures
$H^a$ is
$d$-closed and thus $d$-exact,
\bb P(H) = dQ(H,B)
\ee where $Q$ is linear in the forms $B^a$,
\bb 
\setlength{\fboxsep}{7pt}
\framebox{$Q = R_a(H^b) B^a,$}
\setlength{\fboxsep}{3pt}\label{79}
\ee One may in fact assume that $Q$ involves only the
potentials
$B^a$ of the curvatures of smaller form-degree occuring in
$P$.  To searched-for solution of the Wess-Zumino
consistency condition of which $P$ is the obstruction to a
further lift is of course just $Q$. These are the familiar
Chern-Simons terms, which exist provided one can match the
spacetime dimension $n$ with a polynomial in the curvatures
$H^a$ and the forms $B^a$, linear in $B^a$.

The whole descent associated with $Q$ is generated through
the "Russian formula" \cite{descent,Baulieu1}
\begin{eqnarray} P &=& \tilde{\gamma} Q(H, \tilde{B})
\label{descentCS} \\
\tilde{\gamma} &=& d + \gamma \\
\tilde{B}^a &=& B^{a(p_a,0)} + B^{a(p_a-1,1)} + \cdots
B^{a(0,p_a)}
\end{eqnarray} which follows from the ``horizontability
condition"
\cite{Baulieu1}
\bb
\tilde{\gamma} \tilde{B}^a = H^a.
\ee By expanding (\ref{descentCS}) according to the ghost
number, one gets the whole tower of descent equations.  The
bottom takes the form
$R_a(H^b) B^{a(0,p_a)}$ and is linear in the last ghosts of
ghosts associated with the forms of smaller form degree
involved in $P$. That the bottoms should take this form
could have been anticipated since these are the only
bottoms with the right degrees that can be lifted all the
way to form-degree $n$.  The non-triviality of the bottom
implies also the non-triviality of the whole tower.

It is rather obvious that the Chern-Simons terms are
solutions of the Wess-Zumino consistency condition.  The
main result here is that these are the only solutions that
descend non trivially (up to solutions of type A).

\subsubsection{Chapline-Manton model 1} There is in this
case no non trivial solution of type B since there is no
non trivial descent.

\subsubsection{Chapline-Manton model 2} One may proceed as
for the free theory.  The polynomial
$P$ must be taken in the invariant cohomology of $d$ and so
is a polynomial in the curvatures $F$ with
$F^{r+1}$ identified with zero.  This leads, as in the free
theory, to the Chern-Simons terms 
\begin{equation}
\setlength{\fboxsep}{7pt}
\framebox{$a=F^m A$,}
\setlength{\fboxsep}{3pt}\label{714}
\end{equation}
except that
$F^r A$ is now absent because it can be brought to class A
up to exact terms.  These Chern-Simons terms are available
in all odd dimensions.

\subsubsection{Chapline-Manton model 3}
\label{above}

In this case, the obstruction $P$ is a polynomial in the
improved field strength $G_M$.  One has $P=dQ(G_M, C_M)$
where 
\begin{equation}
\setlength{\fboxsep}{7pt}
\framebox{$Q(G_M, C_M) = R(G_M) C_M$}
\setlength{\fboxsep}{3pt}\label{715}
\end{equation}
is linear in the improved
potential $C_M = C-AB-\frac{1}{2} A dA$.  The Chern-Simons
solution $Q$ exists only in spacetime dimension $4k -1$. As
in the free case, the whole descent associated with $Q$ is
generated through the Russian formula:
\begin{eqnarray} P &=& \tilde{\gamma} Q(G_M, \tilde{C}_M)
\label{descentCS2} \\
\tilde{\gamma} &=& d + \gamma \\
\tilde{C}_M &=& C_M + E_M + L_M + {\tilde C}^{(0,3)},
\end{eqnarray} with $E_M= C^{(2,1)}-\frac12
A^{(1,0)}B^{(1,1)}-\frac12dA A^{(0,1)}-BA^{(0,1)}$ and $L_M
= C^{(1,2)}-\frac12 A B^{(0,2)}-\frac12
A^{(0,1)}B^{(1,1)}$. This follows from the
``horizontability condition"
\bb
\tilde{\gamma} \tilde{C}_M = G_M.
\ee

\subsubsection{Chapline-Manton model 4} Again, one finds as
solutions the familiar {\em higher order Yang-Mills
Chern-Simons not involving $tr F^2$ or
$\omega_3$}. These are available in odd dimensions $>3$.

\subsection{Anomalies of type B}

The anomalies $a$ of type B can themselves be of two types.
They can arise from an obstruction that lives one dimension
higher or from an obstruction that lives two dimensions
higher.  In the first case, the obstruction
$da$ has form degree $n+1$ and ghost number $1$.  This case
is only possible for the free models with $1$-forms and the
second Chapline-Manton model, since there is no
$\gamma$-cohomology in ghost number one for the other
models.  In the other case, the anomaly can be lifted once,
$da + \gamma b =0$.  The obstruction
$db$ to a further lift is then a $(n+2)$-form of ghost
number
$0$.

In the first case, the obstruction $da$ reads
\bb da + \gamma(\hbox{something}) = P_A(H) B^{A(0,1)}
\label{obstruction1}
\ee (we consider explicitly the free case, the second CM
model being handled similarly).  The right-hand side of
(\ref{obstruction1}) is necessarily the $d_1$ of something.
Indeed, it cannot be the $d_k$ ($k>1$) of something, say
$m$, since this would make $m$ trivial: the first
obstruction to the lift of $m$ would have to vanish and $m$
involves explicitly the variables of the $1$-form sector
(see theorem {\bf \ref{triviality}} above). This implies
\bb P_A(H) B^{A(0,1)} = C_{AB}(H) H^A B^{B(0,1)}, \; \;
C_{AB}(H) = - C_{BA}(H)
\label{obstruction2}
\ee so that
$P_A(H) B^{A(0,1)} = d_1(\frac{1}{2}
C_{AB}(H)B^{A(0,1)}B^{B(0,1)})$. One thus needs at least
two $1$-forms to construct such solutions. If
$C_{AB}(H)$ involves the curvatures $H^A$ of the $1$-forms,
it must be such that (\ref{obstruction2}) is not zero.  
The anomaly following from (\ref{obstruction1}) is
\bb 
\setlength{\fboxsep}{7pt}
\framebox{$a = C_{AB}(H) B^{A (1,0)} B^{B(0,1)}$,}
\setlength{\fboxsep}{3pt}\label{722}
\ee and the associated descent is generated through
\bb C_{AB}(H) H^A B^{B(0,1)} = \tilde{\gamma} (\frac{1}{2}
C_{AB}(H)
\tilde{B^A} \tilde{B^B})
\ee

In the second case and for the free theory, the obstruction
$P\in H^{inv}(d)$ is a polynomial in $H^a$ of form-degree
$n+2$, which can be written $P = dQ$ where $Q$ is linear in
the potentials associated with the curvatures of lowest
degree occuring in $P$.  The solution $a$ and the descent
are obtained from the Russian formula (\ref{descentCS}),
exactly as for the counterterms,
\bb 
\setlength{\fboxsep}{7pt}
\framebox{$a = R_a(H^b) C^a.$}
\setlength{\fboxsep}{3pt}\label{724}
\ee  They are linear in the
ghosts and exist only if there are forms of degree $>1$
which are the only ones that can occur in $P$ since
otherwise $a$ is either trivial or of type A. Indeed, if
variables from the
$1$-form sector occur in
$P$, then $P = d_1 a$ (if $P$ is non trivial) and the
descent has only two steps.  But this means that $a$ is the
bottom of the descent and is really of type A.

There is no solution of this type for the Chapline-Manton
model 1 because of the lack of a non-trivial descent. 

For the Chapline-Manton model 2, there is again no anomaly
that can be lifted once because there is no element
belonging to an
$F_k$ yielding upon lifting the appropriate obstruction.

For the Chapline-Manton model 3, solutions descending from
polynomials
$P(G_M)$ in two dimensions higher exist only
in spacetime dimensions equal to
$4k-2$. They are given by 
\bb 
\setlength{\fboxsep}{7pt}
\framebox{$a=Q(G_M)L_M,$}
\setlength{\fboxsep}{3pt}\label{725}
\ee
with $L_M$ defined
in Section \ref{above}. 

Finally, for the Chapline-Manton model 4, one has {\em all
the anomalies of the $SU(N)$ Yang-Mills theory, except
those involving the cocycle $trC^3$ and its lifts} since
they are now trivial.

\section{Conclusions}
\setcounter{equation}{0}
\setcounter{theorem}{0}

In this paper, we have derived the general solution of the
antifield-independ\-ent Wess-Zumino consistency condition
for models involving $p$-forms.  We have justified in
particular why one may assume that the solutions can be
expressed in terms of exterior products of the fields, the
ghosts (which are all exterior forms) and their exterior
derivatives, {\em when these solutions occur in non trivial
descents}.  This is not obvious to begin with since there
are solutions that are not expressible in terms of forms
(those that descend trivially) and justify the usual
calculations made for determining the anomalies.  Once one
knows that the solutions involved in non trivial descents
can be expressed in terms of forms (up to solutions that
descend trivially), one can straightforwardly determine
their explicit form in ghost numbers zero and one.  This was
done in the last section where all counterterms and
anomalies have been classified.  The
counterterms for the free models are either strictly gauge
invariant and given by (\ref{71}) or of the Chern-Simons
type (when available) and given by (\ref{79}). The
counterterms for the Chapline-Manton models are also either
strictly invariant (Eqs. (\ref{72}), (\ref{73}),
(\ref{74}), (\ref{75})) or of the Chern-Simons type. These
Chern-Simons solutions exist for all the models except the
first one in appropriate dimensions (see Eqs. (\ref{714}),
(\ref{715}) respectively and Section {\bf 7.2.5}).

The anomalies may also be either strictly annihilated by
$\gamma$, or lead to a non-trivial descent. The first type
generalizes the anomalies of Dixon and Ramon Medrano
\cite{DixonRM} and exist in the free case and the second CM
model. The more familiar anomalies with a non trivial
descent are analyzed in Section {\bf 7.4} and listed in Eqs.
(\ref{722}), (\ref{724}), (\ref{725}) and below (\ref{725}).

The method applies also
to other values of the ghost number, which are relevant in
the analysis of the antifield-dependent cohomology.

As we have shown in \cite{HKS1}, the natural appearance of
exterior forms holds also for the characteristic
cohomology: all higher order conservation laws are
naturally expressed in terms of exterior products of field
strengths and duals to the field strengths.  It is this
property that makes the gauge symmetry-deforming consistent
interactions for $p$-form gauge fields expressible also in
terms of exterior forms and exterior products
\cite{HK0,HK}.

\section*{Acknowledgements}

The authors are grateful to Glenn Barnich, Friedemann
Brandt, Michel Dubois-Violette,  Christiane Schomblond,
Michel Talon, Claude Viallet and Andr\'e Wilch for useful
discussions. B. K. is ``Aspirant du Fonds National de la
Recherche Scientifique" (Belgium).  MH thanks the Erwin
Schr\"odinger Institute for hospitality while this paper
was completed. This work has been partly supported by
I.I.S.N.-Belgium (convention 4.4505.86) and the ``Actions
de Recherche Concert\'ees" of the ``Direction de la
Recherche Scientifique - ``Communaut\'e francaise de
Belgique".


\begin{thebibliography}{99}
\bibitem{WZ} J. Wess  and B. Zumino, {\em Phys. Lett.} {\bf
37 B} (1971) 95.
\bibitem{CM} G.F. Chapline and N.S. Manton, {\em Phys.
Lett.} {\bf B120} (1983) 105; H. Nicolai and P.K. Townsend,
{\em Phys. Lett.} {\bf 98B} (1981) 257; A. H. Chamseddine,
{\em Nucl. Phys.} {\bf B185} (1981) 403; {\em Phys. Rev.}
{\bf D24} (1981) 3065; E. Bergshoeff, M. de Roo, B. de Wit
and P. van Nieuwenhuizen, {\em Nucl. Phys.} {\bf B195}
(1982) 97.
\bibitem{BRS} C. Becchi, A. Rouet and R. Stora, {\em
Commun. Math. Phys.} {\bf 42} (1975) 127~; {\em Ann. Phys.}
(NY) {\bf 98} (1976) 287.
\bibitem{Zinn} J. Zinn-Justin, {\em Renormalisation of gauge
theories} in {\em Trends in elementary particle theory},
Lecture notes in Physics n$^0$37 (Springer, Berlin 1975); 
{\em Quantum Field Theory and Critical Phenomena}, $3^{\rm
d}$ Edition, Clarendon Press (Oxford: 1995).
\bibitem{BV} I.A. Batalin and G.A. Vilkovisky, {\em Phys.
Lett.} {\bf 102 B} (1981) 27; {\em Phys. Rev.} {\bf D 28}
(1983) 2567.
\bibitem{VT} B. L. Voronov and I. V. Tyutin, {\em Theor.
Math. Phys.} {\bf 50} (1982) 218; {\bf 52} (1982) 628.
\bibitem{ItzZu} C. Itzykson and J.-B. Zuber, ``Quantum
Field Theory", Mc Graw Hill (New York: 1980).
\bibitem{PiguetS} O. Piguet and S. P. Sorella, ``Algebraic
Renormalization", {\it Lecture Notes in Physics} vol. m28,
Springer Verlag (Berlin: 1995).
\bibitem{Weinberg} S. Weinberg, ``The Quantum Theory of
Fields", volume 2, Cambridge University Press (Cambridge:
1996).
\bibitem{BBH2} G. Barnich and M. Henneaux, {\it Phys. Rev.
Lett.} {\bf 72} (1994) 1588; G. Barnich, F. Brandt and M.
Henneaux, {\em Commun. Math. Phys.} {\bf 174} (1995) 93.
\bibitem{HK} M. Henneaux and B. Knaepen, {\em Counterterms
and anomalies for $p$-form gauge theories}, in preparation.
\bibitem{DVTV0} M. Dubois-Violette, M. Talon and C.
Viallet, {\em Phys. Lett.} {\bf 158 B} (1985) 231.
\bibitem{DVTV} M. Dubois-Violette, M. Talon and C. Viallet,
{\em Commun. Math. Phys.} {\bf 102} (1985) 105.
\bibitem{DVTV1} M. Dubois-Violette, M. Talon and C.
Viallet, {\em Ann. Inst. Henri Poincar\'e} {\bf 44} (1986)
103.
\bibitem{MT} M. Talon, ``Algebra of anomalies", Cargese
Summer Inst. Jul 15-31 (1985) 433.
\bibitem{BDK1} F. Brandt, N. Dragon and M. Kreuzer, {\em
Phys. Lett.} {\bf B231} (1989) 263; {\em Nucl. Phys.} {\bf
B332} (1990) 250.
\bibitem{DVHTV} M. Dubois-Violette, M. Henneaux, M. Talon
and C. M. Viallet, {\em Phys. Lett.} {\bf B289} (1992) 361.
\bibitem{Siegel} W. Siegel, {\em Phys. Lett.} {\bf 93 B}
(1980) 170.
\bibitem{BF} I.A. Batalin and E.S. Fradkin, {\em Phys.
Lett.}, {\bf 122B} (1983) 157.
\bibitem{Thierry1} J. Thierry-Mieg, {\em Nucl. Phys.} {\bf
B335} (1990) 334.
\bibitem{Thierry2} L. Baulieu and J. Thierry-Mieg, {\em
Nucl. Phys.} {\bf B228} (1983) 259.
\bibitem{HT} M. Henneaux and C. Teitelboim, {\em
Quantization of Gauge Systems}, Princeton University Press
(Princeton: 1992).
\bibitem{DJT} S. Deser, R. Jackiw and S. Templeton, {\em
Phys. Rev. Lett.} {\bf 48} (1982) 975.
\bibitem{MH} M. Henneaux, {\em Phys. Lett.} {\bf 368B}
(1996) 83.
\bibitem{HK0} M. Henneaux and B. Knaepen, {\em Phys. Rev.}
{\bf D 56} (1997) 6076.
\bibitem{CT} C.\ Teitelboim, {\em Phys.\ Lett.} {\bf 167B}
(1986) 63.
\bibitem{Baulieu} L. Baulieu, {\em On Forms with
Non-Abelian Charges and Their Dualities}, hep-th/9808055.
\bibitem{FT} D.\ Freedman and P.\ K.\ Townsend, {\em Nucl.\
Phys.}\ {\bf B177} (1981) 282.
\bibitem{Dragon} 
F.~Brandt and N.~Dragon, {\em Nonpolynomial gauge invariant interactions of
1-form and 2-form gauge potentials\/}, in {\em Theory of
Elementary Particles}, pp. 149-154, H. Dorn, D. L\"ust, G.
Weigt (eds.) (Wiley-VCH, Weinheim, 1998), hep-th/9709021.
\bibitem{GS} M. B. Green and J. H. Schwarz, {\em Phys.
Lett.} {\bf 149B} (1984) 117.
\bibitem{Baulieu2} L. Baulieu, {\em Phys. Lett.} {\bf 167
B} (1986) 56.
\bibitem{Baulieu1} L. Baulieu, {\em Algebraic construction
of gauge invariant theories}, in {\em Perspectives in
Particles and Fields}, Carg\`ese 1983, M. Levy, J.-L.
Basdevant, D. Speiser, J. Weyers, M. Jacob and R. Gastmans
eds, NATO ASI Series B126, Plenum Press, New York (1983).
\bibitem{XX} L. Romans, {\em Phys. Lett.}  {\bf B 169}
(1986) 374.
\bibitem{XXX} E. Bergshoeff, M. de  Roo, G. Papadopoulos,
M. B. Green and P. K. Townsend, {\em Nucl. Phys.} {\bf B
470} (1996) 113.
\bibitem{HK2} M. Henneaux B.  Knaepen and C. Schomblond,
{\em Lett. Math. Phys.} {\bf 42} (1997) 337.
\bibitem{HKS1} M. Henneaux, B. Knaepen and C. Schomblond,
{\em Commun. Math. Phys.} {\bf 186} (1997) 137.
\bibitem{Sullivan} D. Sullivan, ``Infinitesimal
Computations in Topology", Publ. I.H.E.S. {\bf 47} (1977)
269.
\bibitem{hamermesh} M. Hamermesh, {\it Group Theory},
Addison Wesley (1962).
\bibitem{GarciaK} J. A. Garcia  and B. Knaepen, {\em
Couplings between Generalized Gauge Fields},
hep-th/9807016, to apear in Phys. Lett. B.
\bibitem{Curtright} T. Curtright, {\em Phys. Lett.} {\bf
165B} (1985) 304.
\bibitem{wilch} M. Henneaux and A. Wilch, {\em Phys. Rev.}
{\bf D 58} (1998) 025017.
\bibitem{Saliu} C. Bizdadea and S. O. Saliu, {\em Int. J.
Mod. Phys.} {\bf A 11} (1996) 3523.
\bibitem{Greub} W. Greub, S. Halperin and R. Vanstone, {\em
Connections, curvature and cohomology}, vol. III, Academic
Press (New York: 1976).
\bibitem{Koszul} J. L. Koszul, {\em Bull. Soc. Math. Fr.}
{\bf 78} (1950) 65.
\bibitem{descent} R. Stora, in {\em New Developments in QFT
and Statistical Mechanics}, M. Levy and P. Mitter eds,
Plenum Press (New York: 1977); R. Stora, in {\em Recent
Progress in Gauge Theories}, Lehmann G. et al eds, Plenum
Press (New York: 1984); B. Zumino, in {\em Relativity,
Groups and Topology II}, B. S. De Witt and R. Stora eds,
North Holland (Amsterdam: 1984).
\bibitem{trivialityofd} A.M. Vinogradov, {\em Sov.
Math. Dokl.} {\bf 19} (1978) 1220; M. De Wilde, {\em Lett.
Math. Phys.} {\bf 5} (1981) 351; W.M. Tulczyjew, {\em
Lecture Notes in Math.} {\bf 836} (1980) 22; P. Dedecker
and W.M. Tulczyjew, {\em Lecture Notes in Math.} {\bf 836}
(1980) 498; T. Tsujishita, {\em Osaka J.of Math.} {\bf 19}
(1982) 311; L. Bonora and P. Cotta-Ramusino, {\em Commun.
Math. Phys.} {\bf 87} (1983) 589; P.J. Olver, {\em
Applications of Lie Groups to Differential Equations},
Graduate Texts in Mathematics, volume 107, Springer Verlag
(New York: 1986); R.M. Wald, {\em J. Math. Phys.} {\bf 31}
(1990) 2378; L.A. Dickey, {\em Contemp. Math.} {\bf 132}
(1992) 307. F. Brandt, N. Dragon and M. Kreuzer, {\em Nucl.
Phys.} {\bf B332} (1990) 224; M. Dubois-Violette, M.
Henneaux, M. Talon and C. M. Viallet, {\em Phys. Lett.}
{\bf B267} (1991) 81.
\bibitem{Massey} S. Mac Lane, {\em Homology}, Springer (New
York: 1963). 
\bibitem{Bonora} L. Bonora, P. Cotta-Ramusino, M. Rinaldi
and J. Stasheff, {\em Commun. Math. Phys.} {\bf 112} (1987)
237.
\bibitem{Transgr} H. Cartan, in Colloque de Topologie
(Bruxelles 1950), Masson (Paris: 1951)
\bibitem{DixonRM} J. Dixon and M. Ramon Medrano, {\em Phys.
Rev.} {\bf D 22} (1980) 429.
\end{thebibliography}
\end{document}